\documentclass{elsart}
\usepackage{amssymb}
\usepackage[fleqn]{amsmath}  %%\usepackage{amsmath}
\usepackage{nccmath}
\usepackage{epsfig,epsf,graphicx}
\usepackage
{amsmath}

\usepackage{float}

\def\psla{\rlap \slash}
\def\sla#1{\rlap\slash #1}

\newcommand{\be}{\begin{equation}}
\newcommand{\ee}{\end{equation}}
\newcommand{\cc}{\c c}

\newcommand{\bee}{\begin{eqnarray}}
\newcommand{\eee}{\end{eqnarray}}

\def\be{\begin{eqnarray} &&}

\def\ee{\end{eqnarray}}
\def\bew{\begin{widetext}}
\def\ew{\end{widetext}}

\def\hoje{\number\day\space de \ifcase\month\or
Janeiro,\or Fevereiro,\or Mar\cc o,\or
Abril,\or Maio,\or Junho,\or Julho,
\or Agosto,\or Setembro,\or Outubro,\or Novembro,
\or Dezembro,\fi\space\number\year}

\def\today{\number\day\space de \ifcase\month\or
January,\or February,\or march,\or
April,\or may,\or june,\or July,
\or august,\or September,\or October,\or November,
\or December,\fi\space\number\year}

\begin{document}

\begin{frontmatter} 
\begin{flushright}
    {\large LFTC-18-12/33  \vspace{1cm}}
\end{flushright}
\title{
Unambiguous Extraction of the Electromagnetic Form Factors for Spin-1 Particles on
the Light-Front} 
\author{J. P. B. C. de Melo} \\  
%% \address[LFTC]{
Laborat\'orio de F\'\i sica Te\'orica e Computacional-LFTC \\
Universidade Cruzeiro do Sul, 015060-000, S\~ao Paulo, SP, Brazil
\date{\today}
\maketitle
\vspace{-0.1cm}
\begin{abstract}
The  electromagnetic form factors of a composite vector particle within the 
light-front formulation of the 
Mandelstam formula is investigated. In order to extract the form factors from 
the matrix elements of the plus 
component of the  current in the Drell-Yan frame, where the momentum transfer 
is chosen such that $q^+=q^0+q^3=0$, 
one has in principle the freedom to choose between different linear combinations  
of matrix elements of the current operator. 
The different prescriptions  to calculate the electromagnetic form 
factors, $G_0,G_1$ and $G_2$, i.e.;  charge form factor, 
magnetic and quadrupole respectively. 
If the covariance is respected, all prescriptions give the same results; 
misfortune, is not the situation; the light-front approach 
produce different results, which depend of the 
prescriptions as utilized to extract the electromagnetic form 
factors in the case of the spin-1 particles. 
The main differences of the prescriptions appear because of the 
light-front matrix elements of the electromagnetic current are 
contaminated by the zero-modes  contributions to the 
same with the plus component of the 
matrix elements of the electromagnetic current. 
However, the Inna Grach prescription is immune to the zero-modes 
contributions to the electromagnetic current, then the electromagnetic form factors 
extracted with that prescriptions do not have zero-modes contribution and give the 
same result compared with the instant form 
quantum field theory. Another's prescriptions with the light-front approach are 
contaminated by the 
zero-modes contributions to the matrix elements of the electromagnetic current with the plus component of the current.  
With some relations between the electromagnetic matrix elements of the 
electromagnetic current $J^+_{ji}$, as demonstrated analytical here, it was 
possible to calculate the electromagnetic form factors for spin-1 particles 
without zero-modes or non-valence contributions.
\end{abstract}
\vspace{-0.5cm}
\begin{keyword}
vector particle, electromagnetic form factors, light-front
zero-modes
\end{keyword}
\end{frontmatter}
{\it Introduction:}~The light-front quantum field theory~(LFQFT), is a natural theory 
to describe composite systems, like meson or baryons~\cite{Brodsky98}, and the Fock amplitudes 
of the eigenstates  the LF Hamiltonian reflects the complex structure of the 
hadron obtained with the fundamental interactions
from Quantum Chromodynamics. In respect to the phenomenological success of this 
approach, we should add that it 
produced  results comparable to other approaches for
hadron structure, for example,  Schwinger-Dyson 
methods ~\cite{Maris1999,Roberts1994,Roberts2000}, 
QCD sum rules~\cite{Braguta2004,Braguta22004,Jaus2003,Aliev2004,Savci2009,Braguta2008,Bakulev2009}, 
AdS/QCD frameworks~\cite{Teramond2008}, Effective Field Theory~\cite{Bijnens1998},
and also covariant light-front dynamics~\cite{Leitner2011,Karmanov2007} and recently 
the point-form quantum mechanics~\cite{Elmer2014}. 
On the hand, in LFQFT, the vacuum is trivial and the kinematical group 
contains Lorentz transformations ~\cite{Brodsky98,Perry90}. 
This is an advantage over the usual formalism and allows a great simplification in 
calculations of bound states~\cite {Karmanov2007,Brodsky2004}. 
However, some 
problems with that approach remain, 
the most critical point is the loss of covariance in 
some physical processes~\cite{Pacheco97,Choi2004,Bakker2002,Pacheco99,Naus1999,Li2018}. 
In order to restore the 
full covariance of the electromagnetic current, besides the 
valence component, we need to add the non-valence contributions or zero-modes 
to the matrix elements of the electromagnetic current to keep 
the full covariance~\cite{Pacheco97,Bakker2002,Pacheco99,Naus1999,Pacheco2012}. 
Moreover, the light-front quantum field theory, 
is the natural theory to describe hadronic bound states, 
like pseudoscalar particles~\cite{Leitner2011,Pacheco99,Otoniel2012,Maris97,Fabiano2007,Pacheco2002,Melikhov2002,Yabuzaki2015,Fanelli2016}, 
or spin half particles~\cite{Pacheco2006}. As well, in the last years, some works are dedicate to spin-1 particles, 
with different 
approaches~\cite{Braguta2004,Aliev2004,Savci2009,Karmanov1996,Naus1999,Melikhov2002,Adamuscin2007,Samsonov2003,Grigoryan2007,Aliev2009,Garcia2010,Choi2011,Pitschmann2013,Choi2014,Melo2016,Krutov2016,Krutov2018,Sun2017,Sun2017v2,Pichowsky1999,Hedditch2007,Owen2015,Shultz2015}. 
In addition, some studies of the hadronic properties in the nuclear medium 
was made in the references~\cite{Blunden1996,Blunden1999,Huber1998,Huber2003,Melo2014,Melo2017,deAraujo2018,deMelo2018b}.

{\it The vertex model and electromagnetic current.} The vertex model for the spinor 
structure of the composite  spin-one particle, 
($m_v-q\bar{q}$),  comes from the  model proposed in \cite{Pacheco97}:
\begin{equation}
 \Gamma^{\mu}(k,p)=\gamma^{\mu}-
 \frac{m_v}{2} 
\left(k^{\mu}+k^{\prime\mu}\right)\,D^{-1}_v(k)~,
 \label{vertex}  
\end{equation}
where $m_v$ is the vector spin-1 particle mass, $D_v(k)=(p \cdot k+m_{v} m
- \imath \epsilon)$ and $k^{\prime}=k-p$.

The Mandelstam formula to compute the electromagnetic form factors of the vector particle 
from the plus component of the current,~$J^+=J^0+J^3$,~is given by:
\begin{small}
\begin{multline}
J^+_{ji}=\imath  \int \frac{d^4k}{(2\pi)^4}
 \frac{ Tr\left[\Gamma\Gamma\right]^+_{ji}\,\Lambda(k,p_f)\,\Lambda(k,p_i) }
{((k-p_i)^2 - m^2+\imath\epsilon) 
(k^2 - m^2+\imath \epsilon)((k-p_f)^2 - m^2+\imath \epsilon) }  ,
\label{eq:tria}
\end{multline}
\end{small}
where,~$\gamma^+=\gamma^0+\gamma^3$ is the Dirac matrix, the 
four-vector are $a^\mu=(a^+=a^0+a^3,a^-=a^0-a^3,a_1,a_2)=(a^+,a^-, \vec{a}_{\perp})$, 
and $a.b=\frac{a^+ b^- + a^- b^+}{2}-\vec{a}_\perp \vec{b}_\perp$,~following 
the Light-front formalism~\cite{Brodsky98}. 
The integral for the matrix elements of the current above 
with the Light-front coordinates is~$d^4k=\frac{1}{2}d^2k_{\perp}dk^+dk^-$.

In the equation above, the numerator is given by the Dirac trace:
%% \begin{small} \begin{multline}
\begin{equation}
Tr\left[\Gamma\Gamma\right]^+_{ji}=Tr\left[ 
 \epsilon_j\cdot  \Gamma(k,p_f)
(\sla{k}-\sla{p_f} +m) \gamma^{+}  
(\sla{k}-\sla{p_i}+m) \epsilon_i \cdot \Gamma(k,p_i)(\sla{k}+m) \right] .
\end{equation}
%%  \end{multline} \end{small}
The regularization function in Eq.~(\ref{eq:tria}), is $\Lambda(k,p)=N/[(k-p)^2-m^2_R+ \imath \epsilon]^2
$, which is
chosen to turn the loop integration finite~\cite{Pacheco97}.   
%% In the following/ 12.06.2018 
In the present work, we adopt the  Breit-frame with 
$q^+=q^0+q^3=0$, $q_y=0$ and $q_x\ne 0$ to compute the 
matrix elements of the current. The Cartesian four-vector polarizations 
of the massive vector particle, 
in the instant form representation ($x^\mu=(t,x,y,z)$) in the chosen frame, are given by:
\begin{equation}
\epsilon^{\mu}_{x}=(-\sqrt{\eta},\sqrt{1+\eta},0,0),\,\,
\epsilon_y^{\mu}=(0,0,1,0), \,\,\epsilon_z^{\mu}= (0,0,0,1), \label{polcarti}
\end{equation} 
for the initial state and the final state, 
\begin{equation}
\epsilon^{'\mu}_{x}=(\sqrt{\eta},\sqrt{1+\eta},0,0), \,\,
\epsilon_y^{'\mu}=(0,0,1,0),\,\, \epsilon_z^{'\mu}=(0,0,0,1), \label{polcartf}
\end{equation}  
where $\eta=-\frac{q^2}{4 m^2_{v}}$. 
Given the polarizations vectors, the quantities
 $G_0$, $G_1$ and $G_2$, namely the charge, magnetic and quadrupole form factors
are found as linear combinations of the  matrix elements of the electromagnetic 
current (see e.g. \cite{Pacheco97,Cardarelli1995}). The constraints of covariance, 
parity and current conservation restrict the number of form factors for the 
vector particle to three. From these requirements, the non-vanishing matrix 
elements are  $J^+_{xx}$, $J^+_{yy}$, $J^+_{zz}$  and $J^+_{zx}=-J^+_{xz}$,~therefore 
one relation exists among them, namely the angular condition expressed 
as~\cite{Inna84,Frankfurt93}:
\begin{equation}
\Delta(q^2) = 
 \left( J^+_{yy} - J^+_{zz} \right)
 \left( 1+ \eta \right) = 0 .
\label{eq:angcart}
\end{equation}
If the angular condition is violated, the different   
prescriptions in the literature~\cite{Inna84,Frankfurt93,Chung88,Hiller92}, 
used to obtain the form factors produces different results (see e.g.~\cite{Pacheco97}). 
The source of this problem was traced back to missing zero-mode 
contributions to the matrix elements of 
the current~\cite{Choi2004,Bakker2002,Pacheco2004}. 
In  a recent paper~\cite{Pacheco2012}, 
was analyzed  the contributions of zero modes to the matrix 
elements of plus component of the electromagnetic current,
~$J^+$, coming from  non-vanishing pair production amplitudes (Z-diagrams) in the limit 
of~$q^+\to 0_+$, where it was employed a symmetric form of the vector particle vertex, namely: 
\begin{equation}
\Lambda_s^{\mu} (k,p)=  \Gamma^{\mu}(k,p)\,\Lambda(k,p)+\left[ k\leftrightarrow-k^\prime \right], \label{newsymm}
\end{equation}
which generalizes the vertex function,~Eq.(\ref{vertex}). 
The conclusion of~\cite{Pacheco2012} for the zero-mode contributions
to the matrix elements of the current can be summarized in the following relations:
\begin{equation}
J^{+Z}_{yy}=0, \,\, J^{+Z}_{xx} =
-\eta \  J^{+Z}_{zz}  \,\,\text{and} \,\, J^{+Z}_{zx}  = -\sqrt{\eta} \ J^{+Z}_{zz},  \,\,  
\label{relations}
\end{equation}
where the last two  can be  computed solely from the valence contributions as:
\begin{equation}
J^{+Z}_{zz}=J^{+V}_{yy}-J^{+V}_{zz},
\label{relations1}
\end{equation}
which is a consequence of the angular condition,~Eq. (\ref{eq:angcart}), 
fulfilled by the covariant and current conserving model given by the 
vertex function, Eq. (\ref{newsymm}).  

The final relations for the matrix elements of the plus component of the current, 
can be computed solely in terms of valence matrix elements:
\begin{eqnarray}
J^+_{xx} =  J^{+V}_{xx}-\eta( J^{+V}_{yy}-J^{+V}_{zz})\,\, \text{and} \,\,
J^+_{zx} =  J^{+V}_{zx}-\sqrt{\eta}(J^{+V}_{yy}-J^{+V}_{zz}),
\label{relations2}
\end{eqnarray}
these expressions ensure that  the zero-modes are taken into account for the 
vertex model, Eq. (\ref{newsymm}). The resulting evaluation of the form factors with the 
above matrix elements are in  
agreement with a direct covariant calculation, namely, without resorting 
to the projection onto the light-front. 
The advantage of this strategy is the possibility to apply the relations given in 
the~Eq.~(\ref{relations2}) to compute the form factors of 
composite vector particles for any valence wave function model.

The relations between the matrix elements of the current in the Cartesian 
and, in the light-front spin (helicity) basis,~$I^{+}_{m'm}$~are given 
by~\cite{Pacheco97,Frankfurt93}:
\begin{eqnarray}
& & I^{+}_{11} = \frac{J^{+}_{xx}+(1+\eta) J^{+}_{yy}-
\eta J^{+}_{zz}+2 \sqrt{ \eta} J^{+}_{zx}}{2 (1+\eta)},
\nonumber \\
& & I^{+}_{10} = \frac{\sqrt{2 \eta} J^{+}_{xx}+\sqrt{2 \eta} J^{+}_{zz}
+\sqrt{2} (\eta-1) J^{+}_{zx}}{2(1+\eta)},
\nonumber \\
& & I^{+}_{1-1} = \frac{-J^{+}_{xx}+(1+\eta) J^{+}_{yy}+
\eta J^{+}_{zz}-2 \sqrt{\eta} J^{+}_{zx}}{2 (1+\eta)},
\nonumber \\
& & I^{+}_{00} = \frac{-\eta J^{+}_{xx}+J^{+}_{zz}+2 \sqrt{\eta} J^{+}_{zx}}
{(1+\eta)} \ .
\label{ifront1}
\end{eqnarray}
The elimination of zero-modes for the matrix elements of the current~$I^{+}_{m'm}$ 
through the relations,~Eq's.~(\ref{relations}), (\ref{relations1})~and~(\ref{relations2}), 
leads to the following:
 \begin{equation}
 I^{+Z}_{11} =  0, \, \,  \,  I^{+Z}_{10}=0, \, \,  \,
 I^{+Z}_{1-1} = 0,
 \label{imm}
 \end{equation}
and
 \begin{equation}
  I^{+Z}_{00}=(1+\eta)J^{+Z}_{zz}=(1+\eta)
  \left( J^{+V}_{yy}-J^{+V}_{zz} \right), 
  \label{ang2}
\end{equation}
showing only the~$I^{+Z}_{00}$ component of the  electromagnetic current has  a non-zero 
contribution from the zero-mode~\cite{Pacheco2012}. 
The relation, Eq.~(\ref{ang2}), is also associated with the fulfillment of the angular condition,
\begin{equation}
\Delta(Q^2) = 
(1+2 \eta)
I^{+}_{11}+I^{+}_{1-1} - \sqrt{8 \eta}
I^{+}_{10} - I^{+}_{00}~,
%% = 0~,
\label{ang14}
\end{equation}
where the matrix elements of~$I^{+}_{11}$, $I^{+}_{1-1}$, and~$I^{+}_{10}$, 
due to Eq.~(\ref{imm}), are computed only from the valence terms. 
After the inclusion the zero modes, 
or the non-valence contributions, the angular condition above, results is zero,
$\Delta(Q^2)=0$. These  results was also found in the reference~\cite{Bakker2002} 
for the particular case of~$\gamma^\mu$ vertex coupling for 
the~$q\bar q$ pair to the~$\rho$-meson for a smeared photon-vertex
and further explored in~\cite{Choi2004} for the vector particle coupling 
to the quarks given by Eq.~(\ref{vertex}).

In the next section, we demonstrate for all prescriptions 
utilized in the literature to extract the form factors for spin-1 particles 
and with the plus component of the electromagnetic current,~$J^+$,
~in the Breit-Frame and Drell-Yan condition are equivalent 
to each other if the relations, Eq.~(\ref{relations2}), are used.

{\it Light-Front Prescriptions for the electromagnetic form factors.} 
Therefore, out of the four matrix elements it is possible to combine them
current in different ways. Because this, the linear combinations 
utilized in order to extract the electromagnetic form factors from 
electromagnetic matrix elements of the current~\cite{Pacheco97,Cardarelli1995,Frankfurt93} 
is not unique;~but this leads us to make a
choose which of the current matrix element eliminate, and used to 
the extract the electromagnetic form factors, $G_0,G_1$ and $G_2$, 
then we have different prescriptions to extracted the electromagnetic 
form factors~\cite{Pacheco97,Cardarelli1995}. Following the Ref.~\cite{Pacheco97}, 
the four prescriptions in the literature are written below in the instant form,~(IF), 
basis of spin and with the light-front basis~\cite{Pacheco97,Cardarelli1995}. 

In the reference~\cite{Inna84}, the authors eliminate the $I^{+}_{00}$ component 
of the electromagnetic current in the 
light-front spin basis, and writing in the Cartesian basis also~\cite{Pacheco97}, 
%% is given by:
\begin{eqnarray}
G_0^{GK}& = &\frac{1}{3}[(3-2 \eta) I^{+}_{11}+ 2 \sqrt{2 \eta} I^{+}_{10} 
+  I^{+}_{1-1}]  \nonumber   \\
& = &  \frac{1}{3}[J_{xx}^{+} +(2 -\eta) J_{yy}^{+} 
+ \eta  J_{zz}^{+}], \nonumber \\
G_1^{GK} & = & 2 [I^{+}_{11}-\frac{1}{ \sqrt{2 \eta}} I^{+}_{10}]
=J_{yy}^{+} -  J_{zz}^{+} - \frac{J_{zx}^{+}}{\sqrt{\eta}},
\nonumber \\ 
G_2^{GK}&=&\frac{2 \sqrt{2}}{3}[
\sqrt{2 \eta} I^{+}_{10} - \eta I^{+}_{11}-  I^{+}_{1-1}] 
=  \frac{\sqrt{2}}{3}[J_{xx}^{+}-(1+\eta) J_{yy}^{+} 
+ \eta  J_{zz}^{+}]~.
\end{eqnarray}
With the relations given by the Eq.(\ref{relations}) 
and Eq.(\ref{relations2}), and after made the substitution in expressions for 
the electromagnetic form factors below~\cite{Inna84}:
\begin{eqnarray}
G^{GK \ (+Z) }_{0} & = & 
\frac{1}{3} \left[ J_{xx}^{(+ Z)} 
+ \eta  J_{zz}^{+ Z } \right] =  
 \frac{1}{3} \left[ - \eta J_{zz}^{+Z} 
+ \eta  J_{zz}^{+Z}  \right] = 0,  
\nonumber \\
G_1^{GK~(+Z)} & = &  
\left[
-J_{zz}^{+Z}[gg] - \frac{ J_{zx}^{+Z }} {\sqrt{\eta}} \right] = 
-J_{zz}^{+Z} + 
\sqrt{\eta} \frac{J_{zz}^{+Z} } {\sqrt{\eta}} = 0, 
\nonumber \\
G_2^{GK \ (+Z)} & = & 
\frac{\sqrt{2}}{3} \left( J_{xx}^{+Z} + 
\eta  J_{zz}^{+Z } \right) =   \frac{\sqrt{2}}{3} \left[ -\eta 
J_{zz}^{+ } + \eta  J_{zz}^{+ Z } \right] = 0 ~,
\end{eqnarray}
the zero modes contribution are cancel out, and the electromagnetic 
form factors with that prescription are 
free of non-valence contributions~\cite{Pacheco2012}. 
The results obtained above, show exactly the prescription used  by Grach et al.~\cite{Inna84}, 
have the same results when compared with the usual 
covariant impulse \linebreak 
approximation~\cite{Pacheco97,Pacheco2012}.
%%% instant form approach~\cite{Pacheco97,Pacheco2012}.

%% CCKP 
The authors of the Ref.~\cite{Chung88}, have 
the expressions below for the electromagnetic form factors; 
and, after used the relations given by the Eq's.(\ref{relations}) and (\ref{relations2}), 
the same expression for the electromagnetic form factors given by 
Grach et al.~\cite{Inna84} are obtained, 
%% \vspace{-1.5cm}
\begin{eqnarray}
 G_0^{CCKP} & = & \frac{1}{3 (1 + \eta)}  \biggl[ (\frac{3}{2}-\eta)(I^+_{11}+ I^+_{00}) +
5 \sqrt{2 \eta} I^+_{10} + (2\eta- \frac{1}{2}) I^+_{1-1}  \biggr] \nonumber  \\ 
& = & \frac{1}{6}[2 J_{xx}^{+} + J_{yy}^{+} + 3 J_{zz}^{+}]=
\frac{1}{3}[J^{+V}_{xx} + (2 - \eta )J^{+V}_{yy} + \eta J^{+V}_{zz}]=G_{0}^{GK}~, \nonumber \\ 
G_1^{CCKP}&=&
\frac{1}{ (1 + \eta)} \biggr[I^+_{11}+ I^+_{00} - 
I^+_{1-1} - \frac{2 (1-\eta)}{\sqrt{2\eta}} I^+_{10}\biggl] 
= 
-\frac{J_{zx}^{+}}{\sqrt{\eta}} =  \nonumber  \\
& =& \biggr[ J^{+V}_{yy}-J^{+V}_{zz} -\frac{J^{+V}_{zx}}{\sqrt{\eta}}
\biggl]= G_1^{GK}, \nonumber  \\
G_2^{CCKP} & = &
\frac{\sqrt{2}}{3 (1 + \eta)}  \biggl[  -\eta I^+_{11} -\eta I^+_{00}  +  2 \sqrt{2 \eta}I^+_{10} 
 - (\eta + 2) I^+_{1-1} \biggr]  =  \nonumber \\ 
& =&
\frac{\sqrt{2}}{3} [J_{xx}^{+}-J_{yy}^{+}] = 
\frac{\sqrt{2}}{3}[ J^{+V}_{xx}-(1+\eta )J^{+V}_{yy} + \eta J^{+V}_{zz}]= 
~G_2^{GK}.  
\label{polizou}
\end{eqnarray} 
The electromagnetic form factors given in the reference~\cite{Chung88}, 
after the use of the relations~Eq.~(\ref{relations2}), for the 
matrix elements of the electromagnetic current produced the same results 
if compared with the covariant impulse 
\linebreak approximation calculations 
and the electromagnetic form factors expressions in the Ref.~\cite{Inna84}.
%% %%  if compared with the instant form calculations 
%% and the electromagnetic form factors expressions in the Ref.~\cite{Inna84}.

%% {\it  J. Hiller and Brodsky, S.}
Another prescription utilized in order to obtain 
the electromagnetic form factors in the literature, 
is the Brodsky and Hiller prescription,~(BH)~\cite{Hiller92}. 
After the substitution of the Eqs.~(\ref{relations}) 
and Eq.~({\ref{relations2}), 
in the original Brodsky and Hiller prescription, 
we obtain the following electromagnetic form-factors for spin-1 particle,
\begin{eqnarray}
G_0^{BH} & = & 
\frac{1}{3 (1 + 2 \eta)}  \biggl[ (3 - 2 \eta) I^+_{00}+ 8 \sqrt{2 \eta}I^+_{10} 
+  2 (2 \eta -1 )I^+_{1-1} \biggr]    \nonumber \\ 
& =& \frac{1}{3 (1+2 \eta)}\biggl [
J_{xx}^{+} (1+2 \eta)+ J_{yy}^{+}(2 \eta-1) 
+  J_{zz}^{+}(3+2 \eta) \biggr]\nonumber \\
& = & \frac{1}{3}\biggl[J^{+V}_{xx} + (2-\eta )J^{+V}_{yy} + \eta J^{+V}_{zz}
\biggr] = G^{GK}_{0},
\nonumber \\
G_1^{BH}&=& 
\frac{2}{(1 + 2 \eta)} \biggl[ I^+_{00} - I^+_{1-1} 
+ \frac{(2 \eta -1 )}{\sqrt{2 \eta}} I^+_{10} \biggr] \nonumber  \\
& = &  \frac{1}{(1+2 \eta)} \biggl[- \frac{J_{zx}^{+}}{\sqrt{\eta}}
 (1+2 \eta)- J_{yy}^{+} +  J_{zz}^{+} \biggr]
 \nonumber \\
 & = & [J^{+V}_{yy}- \frac{J^{+V}_{zx}}{\sqrt{\eta}} - J^{+V}_{zz}] = 
G^{GK}_1,
\nonumber \\
G_2^{BH}& = &
\frac{\sqrt{2}}{3 (1 + 2 \eta)} \biggl[ \sqrt{2 \eta }I^+_{10} -\eta I^+_{00} 
-( \eta + 1 ) I^+_{1-1} \biggr]  
\nonumber \\ 
& = &  
\frac{ \sqrt{2}}{3 (1+2 \eta)}
\biggl[ J_{xx}^{+} (1+2 \eta)- J_{yy}^{+}(1+ \eta) - \eta J_{zz}^{+} 
\biggr] \nonumber \\
& = & \frac{\sqrt{2}}{3}[ J^{+V}_{xx}-(1+\eta )J^{+V}_{yy} + \eta J^{+V}_{zz}]
= G_2^{GK}. 
\label{hiller}  
\end{eqnarray} 
With the relations given by~Eqs.~(\ref{relations}) and (\ref{relations2}), 
the final expressions for the electromagnetic form factor 
for the prescription in the Ref.~\cite{Hiller92}, 
given the same expressions as Grach et al.~\cite{Inna84}, and is also free of the zero modes or 
non-valence contributions \cite{Choi2004,Pacheco2012}. 

%% Karmanov
In the reference~\cite{Karmanov1996}, the author use the 
expression for the spin-1 electromagnetic form factor given below, and, after the 
use the relations~Eqs.~(\ref{relations2}),
we obtain again, the exact expressions given in the reference~\cite{Inna84}:
\begin{eqnarray}
G_0^{KA} & = & 
\frac{1}{3}   \biggl[ 2(1 - \eta) I^+_{11} + 4 \sqrt{2 \eta}I^+_{10} 
+  I^+_{00} \biggr] 
\nonumber \\ 
& =& \frac{1}{3}  [ J^+_{xx} + J^{+}_{yy} ( 1 -2 \eta) + (2 \eta +1) J^+_{zz} ]   \nonumber \\
& = & \frac{1}{3}\biggl[J^{+V}_{xx} + (2-\eta )J^{+V}_{yy} + \eta J^{+V}_{zz}
\biggr] = G^{GK}_{0},
\nonumber \\
G_1^{KA} & = & \biggl[ 2 I^{+}_{11}  - \sqrt{\frac{2}{\eta}}  I^{+}_{10} \biggr] 
~= ~  \biggl[J^{+}_{yy}  - \frac{J^+_{zx}}{\sqrt{\eta}} - J_{zz}^{+} \biggr]
 \nonumber \\
 & = & [J^{+V}_{yy}- \frac{J^{+V}_{zx}}{\sqrt{\eta}} - J^{+V}_{zz}] = 
G^{GK}_1,
\nonumber \\
G_2^{KA}& = &
\frac{ 2 \sqrt{2}}{3} \biggl[ (1 + \eta) I^+_{11} - \sqrt{2 \eta} I^+_{10} -I^+_{00
} \biggr]  
\nonumber \\ 
& = &  
\frac{ \sqrt{2}}{3}
\biggl[ J_{xx}^{+} +  (1+ \eta) J^+_{yy} -(2+\eta) J_{zz}^{+}  
\biggr] \nonumber \\
& = & \frac{\sqrt{2}}{3}[ J^{+V}_{xx}-(1+\eta )J^{+V}_{yy} + \eta J^{+V}_{zz}]
= G_2^{GK} . 
\label{Karmanov}  %% line 523 
\end{eqnarray} 
Finally, in the case of reference~\cite{Frankfurt93}, 
the replacement of the relations,~Eq.(\ref{relations}), and Eq.~(\ref{relations2}), give 
identical expression for the electromagnetic form factors for spin-1 particles, 
%%  by Inna Grach et al.:
\begin{eqnarray}
G_0^{FFS}& = & 
\frac{1}{3 (1 +  \eta)}  \biggl[ (2 \eta +3 ) I^+_{11} 
+ 2 \sqrt{2 \eta} I^+_{10} - \eta I^+_{00} + 
(2 \eta + 1) I^+_{1-1} 
\biggr]  \nonumber \\
%% &= & 
%% \frac{1}{(1+2 \eta)} \biggl[ \frac{J_{zx}^{+}}{\sqrt{\eta}} (1+2 \eta)- J_{yy}^{+} +  J_{zz}^{+} \biggr]
%%  \nonumber \\
& = &  \frac{1}{3}[ J_{xx}^{+} + 2 J_{yy}^{+} ]  
\nonumber \\
& = &  
\frac{1}{3}[J^{+V}_{xx}+(2 - \eta )J^{+V}_{yy} + \eta J^{+V}_{zz}] = G_0^{GK}~, 
\nonumber \\
G_1^{FFS}&  = &  G_1^{CCKP} = G_1^{GK},   \nonumber \\
G_2^{FFS} & = & G_2^{CCKP}= G_2^{GK}~, 
\label{frederico}
\end{eqnarray}  
compared with the expressions obtained by Inna Grach et.~\cite{Inna84}.

In order to see the broken of the covariance for the prescriptions 
present here, the differences between that prescriptions 
are given. For example, for Inna Grach et al.~\cite{Inna84}~(GK) and 
Brodsky and Hiller~\cite{Hiller92}~(BH), 
\begin{eqnarray}
 \delta [G^{BH}_0 - G^{GK}_0 ] & = &  -\frac{(3 - 2 \eta )}{3 (1 + 2 \eta )} 
 \left[
(1+2 \eta)
I^{+}_{11}+I^{+}_{1-1} - \sqrt{8 \eta}
I^{+}_{10} - I^{+}_{00} \right] \nonumber \\
& = & -\frac{(3 - 2 \eta) }{3 (1 + 2 \eta )}   \Delta(Q^2), \nonumber \\ 
 \delta [G^{BH}_1 - G^{GK}_1] & = &  
 -\frac{2}{(1+ 2 \eta) } \left[
(1+2 \eta)
I^{+}_{11}+I^{+}_{1-1} - \sqrt{8 \eta}
I^{+}_{10} - I^{+}_{00} \right] \nonumber \\
& = &  -\frac{2 }{(1 + 2 \eta )} \Delta(Q^2), \nonumber \\ 
\delta [G^{BH}_2 - G^{GK}_2] & = &  
\frac{ 2 \sqrt{2} \eta}{3 (1+ 2 \eta) } \left[
(1+2 \eta)
I^{+}_{11}+I^{+}_{1-1} - \sqrt{8 \eta}
I^{+}_{10} - I^{+}_{00} \right] \nonumber \\
& = &  \frac{2 \sqrt{2} \eta  }{3 (1 + 2 \eta )} \Delta(Q^2)~.
\label{difbhgk}
\end{eqnarray}
And the differences for the  electromagnetic form factors, from the 
prescription utilized by Chung et al.,~\cite{Chung88}~(CCKP), and Inna Grach, 
are given by, 
\begin{eqnarray}
 \delta [G^{CCKP}_0 - G^{GK}_0 ] & = &  -\frac{(3 - 2 \eta )}{6 (1 +  \eta )} 
 \left[
(1+2 \eta)
I^{+}_{11}+I^{+}_{1-1} - \sqrt{8 \eta}
I^{+}_{10} - I^{+}_{00} \right] \nonumber \\
& = & -\frac{(3 - 2 \eta) }{6 (1 +  \eta )}   \Delta(Q^2)~, \nonumber \\ 
 \delta [G^{CCKP}_1 - G^{GK}_1] & = &  
 -\frac{1}{(1+\eta) } \left[
(1+2 \eta)
I^{+}_{11}+I^{+}_{1-1} - \sqrt{8 \eta}
I^{+}_{10} - I^{+}_{00} \right] \nonumber \\
& = &  -\frac{1 }{(1 + \eta )} \Delta(Q^2)~, \nonumber \\ 
\delta [G^{CCKP}_2 - G^{GK}_2] & = &  
\frac{ \sqrt{2} \eta}{3 (1+  \eta) } \left[
(1+2 \eta)
I^{+}_{11}+I^{+}_{1-1} - \sqrt{8 \eta}
I^{+}_{10} - I^{+}_{00} \right] \nonumber \\
& = &  \frac{ \sqrt{2} \eta  }{3 (1 +  \eta )} \Delta(Q^2)~.
\label{difcckpgk}
\end{eqnarray}
For the prescriptions given by the author in the 
reference~\cite{Karmanov1996}, the differences between the electromagnetic form factors are,
\begin{eqnarray}
 \delta [G^{KA}_0 - G^{GK}_0 ] & = &  \frac{1}{3} 
 \left[ I^{+}_{00}  - I^{+}_{1-1} - 2 \sqrt{2  \eta}I^{+}_{10} 
 - I^+_{11} (1 + 2 \eta ) \right] 
 \nonumber \\
& = & -\frac{\Delta(Q^2)}{3},
\nonumber \\ 
 \delta [G^{KA}_1 - G^{GK}_1] & = &  0,
 \nonumber \\ 
\delta [G^{KA}_2 - G^{GK}_2] & = &  
\frac{ -2 \sqrt{2} }{3} \left[
I^{+}_{00} - I^{+}_{1-1} +  \sqrt{8 \eta} I^+_{10}
- ( 2 \eta +1) I^+_{11} \right] \nonumber \\
& = &  \frac{ 2 \sqrt{2}  }{3} \Delta(Q^2).
\label{difkagk}
\end{eqnarray}
Also, the case for Frankfurt et al., prescription~\cite{Frankfurt93} and 
Inna Grach~\cite{Inna84} is, 
\begin{eqnarray}
 \delta [G^{FFS}_0 - G^{GK}_0 ] & = & \frac{\eta }{3 (1 +  \eta )} 
 \left[
(1+2 \eta)
I^{+}_{11}+I^{+}_{1-1} - \sqrt{8 \eta}
I^{+}_{10} - I^{+}_{00} \right] \nonumber \\
& = & -\frac{\eta }{3 (1 +  \eta )}   \Delta(Q^2)~, \nonumber \\ 
 \delta [G^{FFS}_1 - G^{GK}_1] & = &  
 -\frac{1}{(1+\eta) } \left[
(1+2 \eta) I^{+}_{11}+I^{+}_{1-1} - \sqrt{8 \eta}
I^{+}_{10} - I^{+}_{00} \right] \nonumber \\
& = &  -\frac{1 }{(1 + \eta )} \Delta(Q^2)~, \nonumber \\ 
\delta [G^{FFS}_2 - G^{GK}_2] & = &  
\frac{ \sqrt{2} \eta}{3 (1+  \eta) } \left[
(1+2 \eta) I^{+}_{11}+I^{+}_{1-1} - \sqrt{8 \eta}
I^{+}_{10} - I^{+}_{00} \right] \nonumber \\
& = &  \frac{ \sqrt{2} \eta  }{3 (1 +  \eta )} \Delta(Q^2)~.
\label{diffsgk}
\end{eqnarray}
The  differences between the electromagnetic form factors with the prescriptions present 
here,~Eq.~(\ref{difbhgk}),~Eq.~(\ref{difcckpgk}),~Eq.~(\ref{difkagk}) and Eq.(\ref{diffsgk}). 
are proportional to the angular condition,~Eq.(\ref{ang14}),
$\delta G_i\propto \Delta(Q^2)$, 
here,~$i=0,1,2$,~for charge, magnetic and quadrupole electromagnetic form factors. 
In the case of $G^{KA}_1$, the magnetic form factor, from ref.~\cite{Karmanov1996}, is exact the 
same given in the reference~\cite{Inna84}, and, that difference, is 
zero,(see in ~Eq.~(\ref{difkagk})). In some sense, the differences between the prescriptions 
to extract the electromagnetic form factors, give the measure of the broken the rotational 
symmetry in the light-front approach. However, like  discussed before, if the non-valence contributions, 
or zero modes is included 
properly, the angular condition expression given zero, and, the results for the  differences in the 
expressions above for the electromagnetic form factors is also zero~(see the Fig.~\ref{diff1} 
and Fig.~\ref{diff2},~left). 
\vspace{0.54cm}
%\texttt{ }
\begin{figure*}[htb]  %% Fig.2
%%\vspace{2.cm}
\includegraphics[scale=0.285]{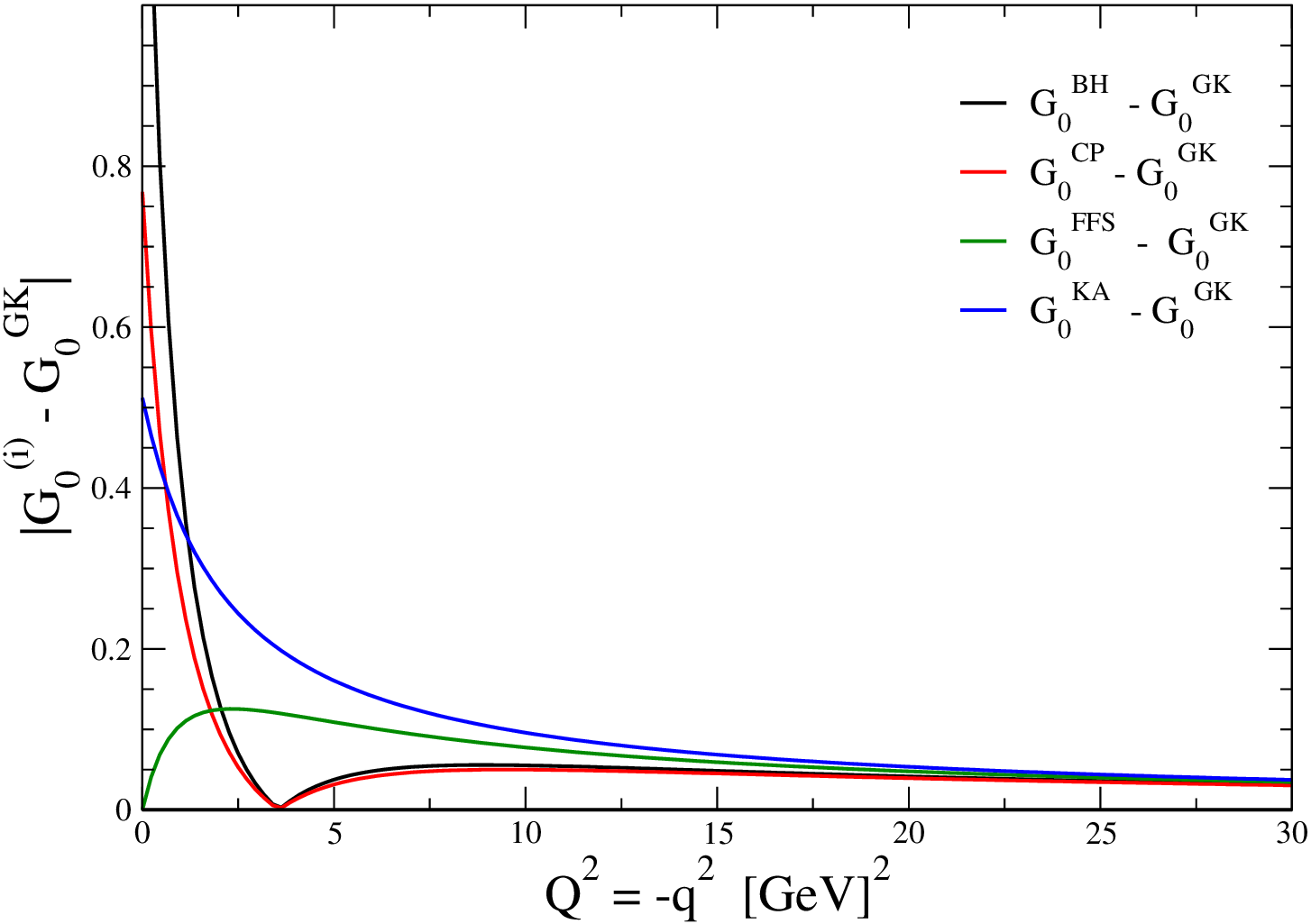}
\includegraphics[scale=0.285]{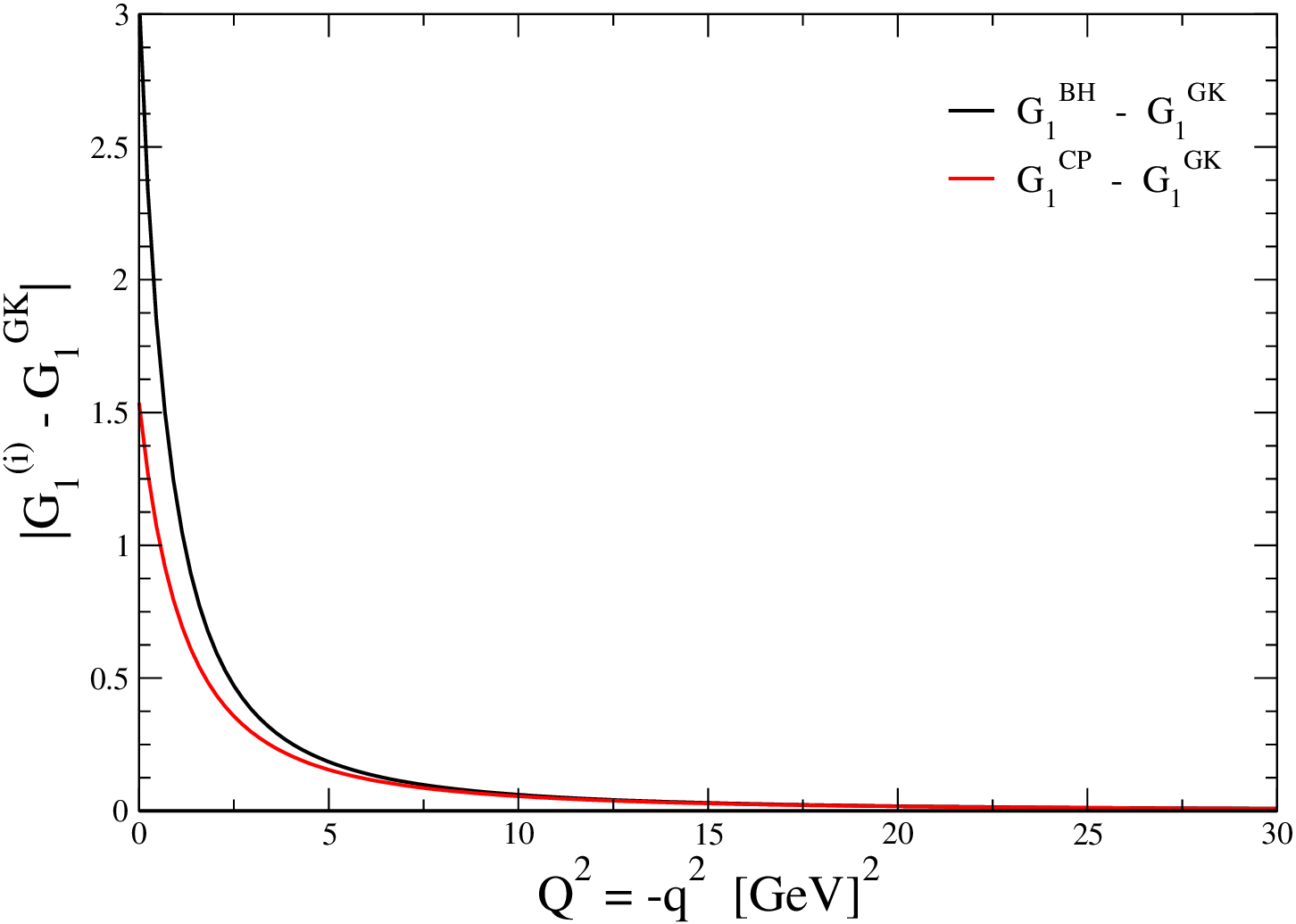}
\caption{The differences among the electromagnetic form factor, $\delta G0$ 
and $\delta G_1$, given by the Inna Grach~\cite{Inna84} 
prescription and other prescriptions in the 
literature~\cite{Karmanov1996}~(KA),\cite{Chung88}~(CP),\cite{Hiller92}~(BH) and 
\cite{Frankfurt93}~(FFS), the Eq.(\ref{difbhgk},\ref{difcckpgk},\ref{difkagk}) and 
Eq's.(\ref{diffsgk}), above. 
Right, the differences for $G_0$ form factors, and left, $G_1$. If the zero mode, or 
non-valence contributions are include, the 
differences for all prescriptions are zero, because the differences are proportional 
to the angular condition,~Eq.(\ref{ang14}).
}
\label{diff1} %% Fig.2
\end{figure*}

\vspace{1.30cm}  
%\texttt{ }
\begin{figure*}[htb]  %% FIG 2
%%\vspace{2.cm}
\includegraphics[scale=0.285]{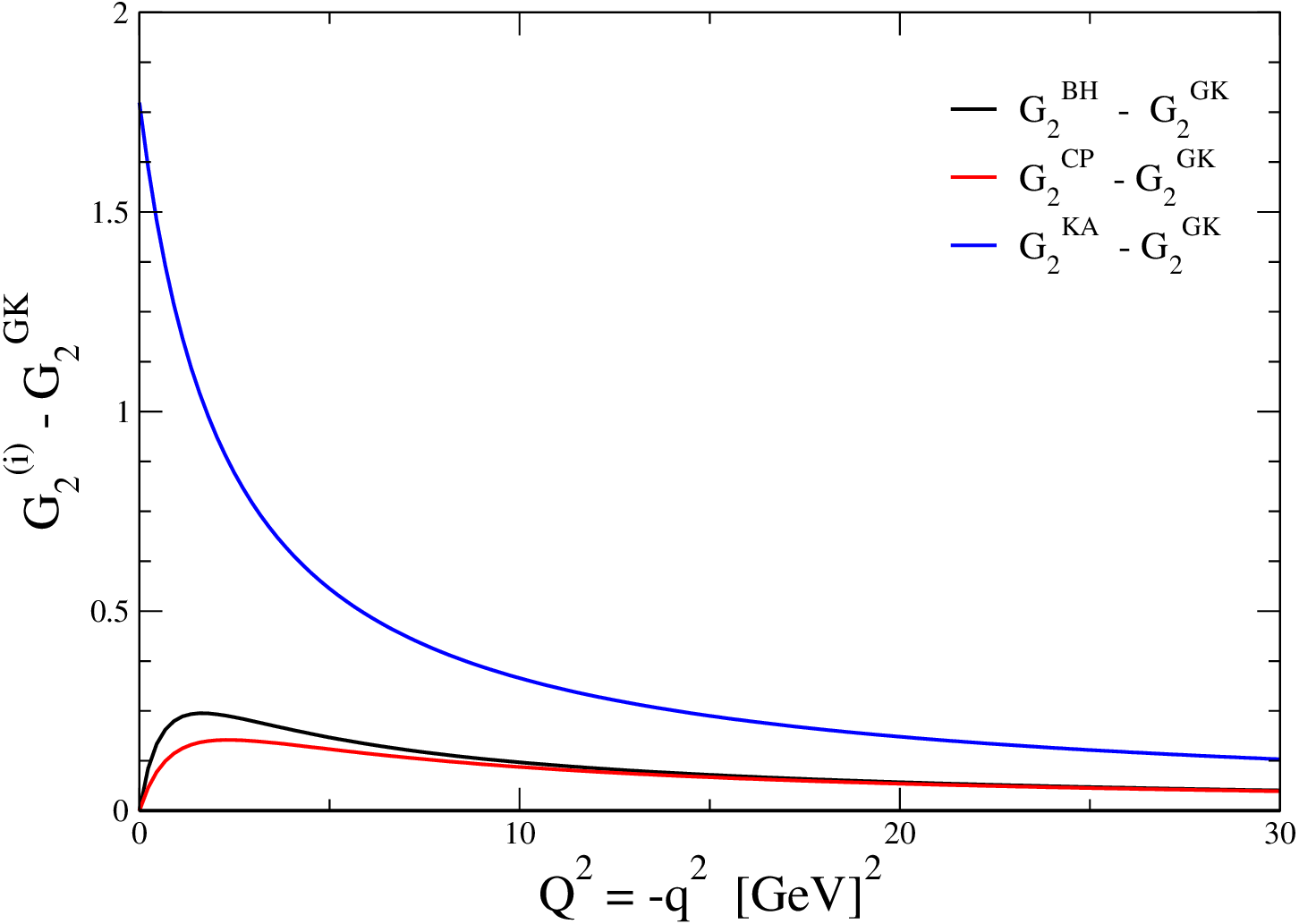}
\caption{The differences for the quadrupole form factor~$\delta G_2$,~for the prescriptions present here 
is show in the figure above. 
The prescriptions given in the reference~\cite{Frankfurt93}, have for the quadrupole rho meson 
form factor, the same expression in the reference~\cite{Chung88}, 
(see the Eqs.~(\ref{difbhgk}),~(\ref{difcckpgk}),~(\ref{difkagk})~and Eqs.~(\ref{diffsgk})).
}
\label{diff2} 
\end{figure*}

\vspace{0.67cm}  %% Fig. 3
\begin{figure*}[htb]
\includegraphics[scale=0.26]{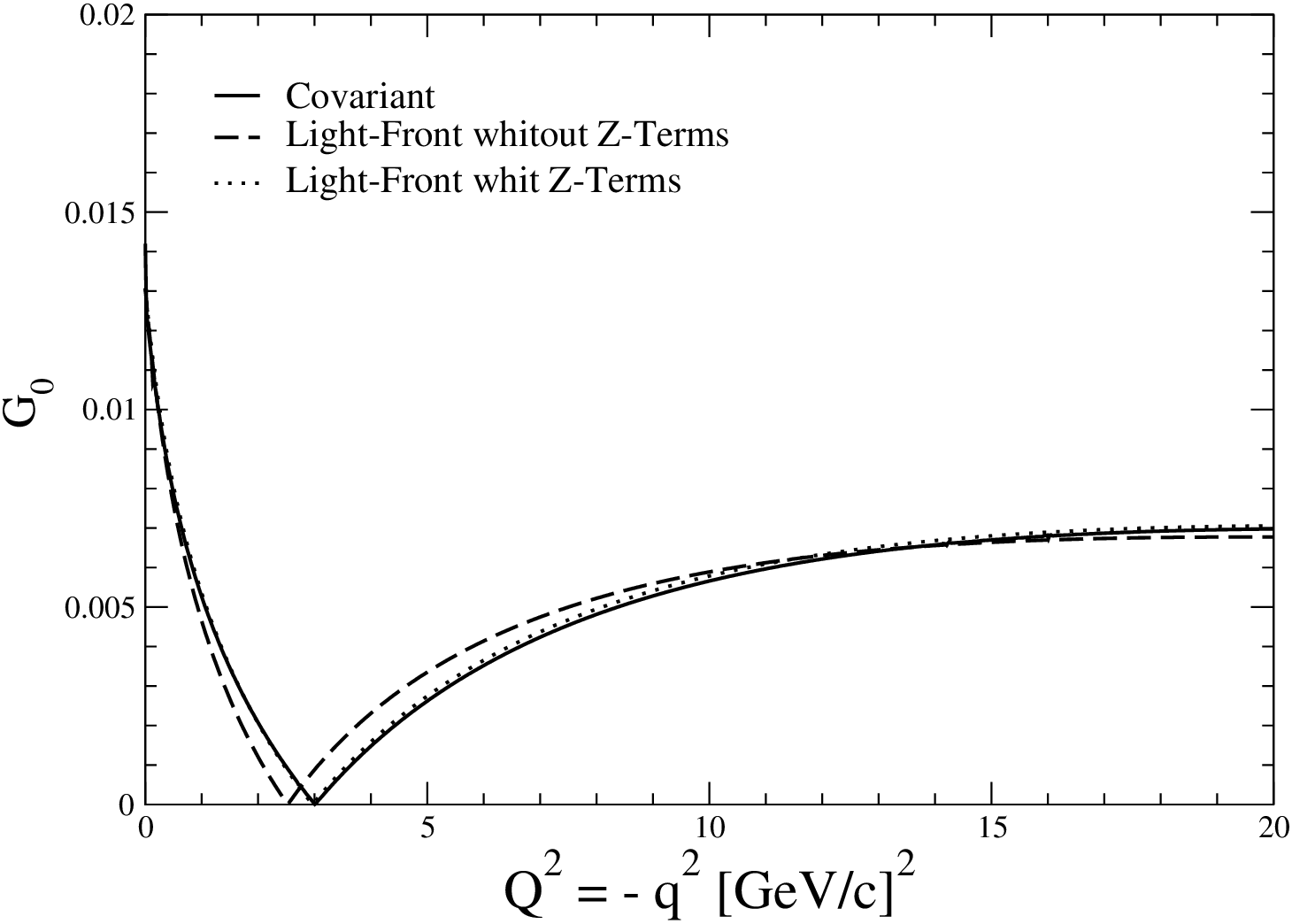}
\hspace{0.1cm}
\includegraphics[scale=0.26]{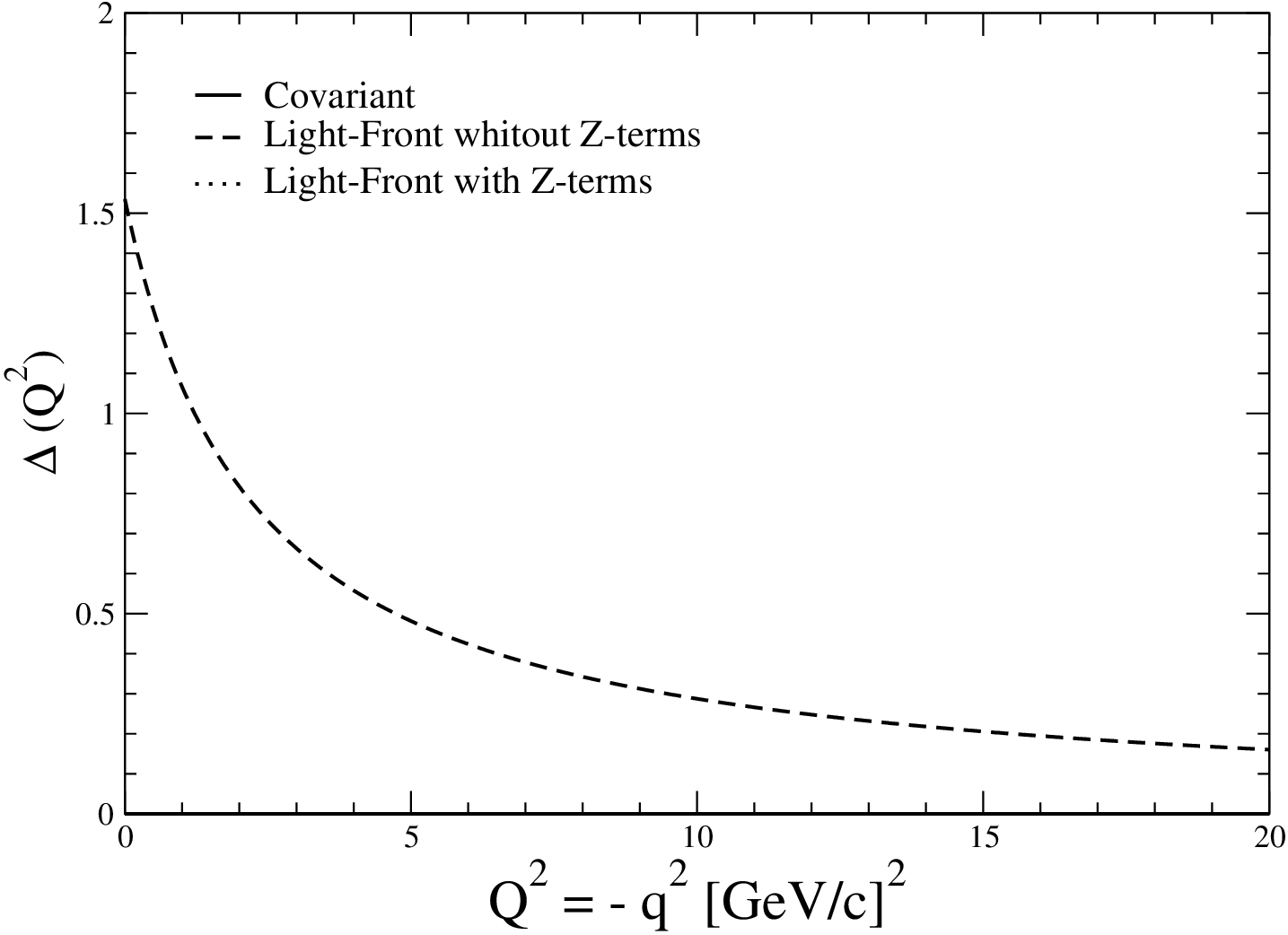}
\caption{(Left),~The zero of the charge form factor with the Eq.~(\ref{sum1})), 
and,~(right),~the angular condition,~$\Delta(Q^2)$,~each figure 
calculated with instant form and light-front approach. For both, calculations, 
the quark mass, is~$m_q=0.430~GeV$ and $m_R=3.0~GeV$.
In the figures above, if the z-terms, or zero modes is not include, the 
value of zero for Eq.~(\ref{sum1}), is different~(about~$2.4~GeV^2$), and also, the angular 
condition is not satisfied. The figure at left, show the angular condition, and, the 
strong breaking of the rotational symmetry, if the zero modes, or pair terms,  
is not taken in account.
}
\label{qzero} 
\end{figure*}

For all prescriptions presented here,~\cite{Inna84,Chung88,Hiller92,Frankfurt93}, 
if the angular condition is satisfied,~Eq.~(\ref{ang2}), follows immediately 
the expression below for the electromagnetic charge form factor,~$G_0$: 
\begin{equation}
 G_0~=~\frac{1}{3}\left[ ~J^+_{xx} + 2 J^+_{yy} \right]~.
 \label{sum1}
\end{equation}
The equation above, produce the same charge electromagnetic 
form factor with the instant form basis calculations, and, 
is exactly the same for all prescriptions given above~\cite{Pacheco97,Cardarelli1995}. 
Though, if the zero modes, or pair terms, are not included, 
the results with the instant form basis and light-front are not the same, and, 
need to add the zero modes or non-valence contributions to have the same 
results for both approaches.  %%%% ,instant form and light-front spin basis calculations.

The charge electromagnetic form factor for spin-1 particles, 
like the deuteron \cite{Garcon2001,Gilman2001}, or 
rho meson~\cite{Pacheco97,Choi2004,Cardarelli1995,Bhagwat2008,Roberts2011}, 
have a zero, i.e., $G_0(q^2_{zero})=0$. In order to see the zero position of the 
charge form factor~($G_0$), with the present light-front model, for the 
expression given above,~Eq.~(\ref{sum1}), we show the Fig.~(\ref{qzero}),~right, 
(this figure, is not normalized to 1), and, the zero for this sum appears when the moment 
transfer is about~$\simeq3.0~GeV^2$, and the inclusion of the zero modes; 
for another side, if the zero modes is not include, the position of this zero in 
the momentum transfer is different,~$\simeq2.4~GeV^2$~(see also the angular condition, 
in the Fig.~(\ref{qzero}),~left). In order to keep the correct position 
of the charge electromagnetic form factor for spin-1 particles, is 
crucial include the non-valence contributions or zero-modes to the 
electromagnetic current for spin-1 particles.

In the discussions above, we have demonstrated, with the use of the relations given in 
the reference~\cite{Pacheco2012}, ie., Eqs.~(\ref{relations}), (\ref{relations1}) and (\ref{relations2}), 
for all prescriptions found in  the literature with the light-front approach for spin-1 particles, 
given the same expression for the electromagnetic form factors, charge, magnetic and quadrupole. 
That procedure, is equivalent to take into account the zero modes, or, non-valence contributions 
to the electromagnetic current for the spin-1 particles~\cite{Pacheco2012}. In the last, the zero-mode, 
are very important to keep the full covariance with the light-front 
approach~(see more in the references \cite{Choi2004,Pacheco99,Pacheco2002,Pacheco2004,Pacheco98}).

{\it Vector decay constant:} Besides the electromagnetic form factors, also the 
electromagnetic decay constant of the rho meson was calculated with the present 
model~\cite{Pacheco97}. We use the following expression to the decay constant 
for spin-1 particles~\cite{Pacheco2006},
\begin{equation}
\imath \epsilon_{\lambda}^{\mu} \sqrt{2} f_{\rho} =
 \langle 0|\bar{q} \gamma^{\mu} q|p\rangle~.
 \end{equation}
The expression for the decay constant with the considered model here, is given by the 
following expression, 
\begin{small}
\begin{eqnarray}
 f_{\rho}=\- -\imath \frac{N_c N}{4 (2\pi)^4} 
\int\frac{ d^2k_{\perp} dk^+ dk^-
Tr\left[(\psla{k}-\psla{p}+m) \gamma^+ (\psla{k}+m )\, \epsilon_z 
\cdot \Gamma\right] \,
\Lambda(k,p)
}
{  k^+(p^+ - k^+) 
(k^--k^-_{on}+\frac{\imath \epsilon}{k^+})
(p^- - k^- -\frac{(p-k)_{\perp}+m^2}{p^+-k^+}+\frac{\imath \epsilon}{p^+-k^+})
}~, \nonumber
\end{eqnarray}
\end{small}
where the polarization vector chosen is ~$\epsilon^+_z=1$ and 
the vector particle it is in the rest frame,~$p^{\mu}=(p^0,\vec{0})$. 
After the Dirac trace performed and the $k^-$ integration done, result in:
  \begin{eqnarray}
   f_\rho =  \dfrac{N_c N}{m_\rho} \int \dfrac{d^2 k_\perp dx }{(2
    \pi)^3} \dfrac{tr[\theta^+]}{x (1 - x)^3 (m^2_\rho - M^2_0 )
     (m^2_\rho - M^{2}_{R}(m_R,m^2))^2 }  \label{cc26} \ ,
\end{eqnarray}
where~$tr[\theta^+]$, is the function defined below from Dirac trace in the expression 
above:
\begin{equation}
tr[\theta^+]  =    \left( - 4 k^{+2}+ 4 k_{\perp}^2 + 4 k^+ P^+ 
 + 4m^2 \right) -    
  \dfrac{m_{\rho}}{2}\dfrac{(2k^+
  - P^+)(k_{on}^- - k^+)4m}{ \left(
  \dfrac{P^+ k_{on}^- + P^- k^+}{2}  + m_{\rho}m \right) }~,
\end{equation}
and with ~$M_R^2(m^2_a,m^2_b)=\frac{k_{\perp}+m_a^2}{x} + 
\frac{k^2_{\perp}+m_b^2}{1-x}$,~ $M_0=M^2(m^2,m^2)$ and $x=k^+/p^+$. The constant $N$, 
is found after the normalization condition for the charge electromagnetic form factor, 
$G_0(0)=1$. 
The function $\theta^+$,~have two terms, a part without $k^-$ dependence and another 
with~$k^-$ dependence.  
In refs.~\cite{Pacheco97,Pacheco99,Naus1999,Pacheco2012}, it was shown for the 
matrix elements of the electromagnetic current~$J_ {ji}^+$, with
terms proportional a $k^-$, can lead to breaking of the rotational symmetry,
but not necessarily; as in the case of the pi meson, where despite 
there being proportionately energy in the light-front,
the contribution of pair terms or zero-modes vanish~\cite{Pacheco99,Otoniel2012}. 
For the case of the present work,~the rho meson decay constant, 
calculated with the vertex,~Eq.~(\ref{vertex});
~the second term, which could make a contribution, does not contribute, because
the structure of the vertex used~ref.~\cite{Pacheco97}; but, for the ref.~\cite{Choi2014}, 
the zero-modes terms survives, in virtue of the vertex structure utilized.

\begin{table}[htb]
\begin{center}
\vspace{0.2cm}
\caption{$\rho$-meson low-energy electromagnetic observables 
calculate with the present light-front model,~and compared with 
diferents models in the literature.
}
\label{tab1} 
 \vspace{0.2cm}
\begin{tabular}{lllll}
\hline 
\hline
                                  &~$f_{\rho}$~[MeV] &~$\mu~[e/2m_{\rho}]$ &~$Q_d~[e/m^2_{\rho}]$&~$<r^2>~[fm^2]$ \\
\hline This work                  &~153.66           &~2.10                &~-0.898           & ~0.267            \\
Pichowsky~\cite{Pichowsky1999}    &~153.95           &~2.69                &~-0.055           &~0.61              \\
Jaus~\cite{Jaus2003}              &~-                &~1.83               &~-0.330              &~~-               \\
Aliev~\cite{Savci2009}            & ~-               &~$2.4\pm0.4$        &~$0.85\pm0.15$       & ~-               \\
Biernat~\cite{Elmer2014}          &  ~-              &~2.20               &~-0.47               & ~-               \\ 
Choi~\cite{Choi2004}              &~ -               &~1.92               &~-0.430               &~-               \\
%%%% De Melo~\cite{Pacheco97}          &  ~-              &~2.19               &~-0.79               & ~0.37       \\
Melikhov~\cite{Melikhov2002}      &  ~-              &~2.35               &~-0.364                  &~-              \\                   
Samsonov~\cite{Samsonov2003}      &~-                &~$2.00\pm0.3$       &~-                   &~-                   \\
Pitschmann~\cite{Pitschmann2013}  &  ~-              &~2.11               &~-0.850              & ~0.26              \\
Krutov~\cite{Krutov2016,Krutov2018} &~152$\pm$8      &~2.16$\pm$0.03       &-~                   &~0.56$\pm$0.04    \\
Sun~\cite{Sun2017,Sun2017v2}       & ~-              &~2.06               &~-0.323              &~0.52              \\
Hawes~~\cite{Pichowsky1999}        &~~-              &~2.69               &~-0.055              &~~0.61             \\
Cardarelli~\cite{Cardarelli1995}   & ~-              &~2.26               &~-0.367              &~0.35          \\
Bhagwat~\cite{Bhagwat2008}        & ~-               &~2.01               &~-0.41                &~0.54        \\
Roberts~\cite{Roberts2011}        &~-                &~2.11               &~-0.85               &~0.31         \\
Serrano~\cite{Serrano2015}        & ~~               &~2.57               &~-1.05               &~0.67         \\
Gudi\~no~\cite{Gudino2015}        &~-                &~2.1$\pm$0,5        &~-                   &~-            \\
Simonis~\cite{Simonis2016}        &~-                &~2.06               &~-                   &~-             \\
Simonis~\cite{Simonis2018}        &~-                &~2.17               &~                   &~              \\
Owen~\cite{Owen2015}              &~                 &~2.145              &~-0.733             &~0.670   \\
Shultz~\cite{Shultz2015}          &~-                &~2.17               &~-0.540             &~~0.30   \\
\hline                                                                                   
PDG~\cite{PDG2018}                   &~153~$\pm$~8      &    -               &    -                &  -            \\
\hline 
\hline
\end{tabular}
\end{center}
\end{table} 
\vspace{0.5cm}

{\it Results.}~The electromagnetic form factors,~$G_0,G_1$ and $G_2$ 
was calculated here with the vertex model~$\rho-q\bar{q}$,~Eq.(\ref{vertex}), utilized 
previously in the reference~\cite{Pacheco97}. 
Was already noted in previous works, the various prescriptions in the literature does not given 
the same results for the electromagnetic form factors,  
compared with the covariant impulse approximation~\cite{Pacheco97,Bakker2002}.
%% instant form calculation~\cite{Pacheco97,Bakker2002}. 
Because the  
dependence of which matrix element of the electromagnetic current is eliminate with the angular 
condition equation,~\cite{Pacheco97,Inna84}, we have some freedom 
in eliminate the matrix elements~$I^+_{m m'}$, but, if the matrix element 
$I^+_{00}$ is 
not eliminate, the rotational symmetry is broken~\cite{Pacheco97,Bakker2002,Pacheco2012}, 
and, we have the zero modes, or, non-valence contributions. 
In the present work, after the use of the relations given in Eq.(\ref{relations2}), all 
prescriptions found in the literature given the same results for the 
electromagnetic form factors~\cite{Pacheco97,Pacheco2012}.  
However, the prescription utilized by Inna Grach et al.,~\cite{Inna84}, give exactly the same 
results, if compared with the covariant impulse approximation, for all 
observables calculated here, ie., the electromagnetic form factors, 
~$G_0,G_1,G_2$, electromagnetic radius and the rho meson decay constant.

With the constituent quark mass $m=m_u=m_d=0.430~GeV$, the regulator mass $m_R=3.0~GeV$ and the 
experimental rho mass,~$m_{\rho}=~0.770~GeV$, we have obtained the value of the decay 
constant~$f_{\rho}=154~MeV$,~very close with the  
experimental data~$153\pm8~MeV$~\cite{PDG2018}; the electromagnetic radius 
is $<r^2>=0.267~fm^2$,~the magnetic moment~$\mu=2.205~[e/2 m_\rho]$,
~and the quadrupole moment~$Q_d=-0.0586~fm^2$, the values of the 
observables obtained here, are comparable  with 
another's  models in the literature~(see the table I). 

In the Fig.\ref{ffactorsg0g1}, 
we present the charge and magnetic form factors,~$G_0$~and~$G_1$, 
calculated with the parameters given above for 
various prescription in the literature.  

The calculations present in the Fig.~\ref{ffactorsg0g1}, 
are made without non-valence  contributions, dashed lines 
and with add the non-valence contributions, solid lines. 
All calculation are compared with the covariant impulse approximation calculation, 
black solid line. It is seen in the figures 
for the electromagnetic form factors, the Inna Grach prescription~\cite{Inna84}, give 
the same result compared with the covariant impulse approximation. After the use of the 
relations Eqs.~(\ref{relations1}) and Eq.~(\ref{relations2}), which correspond in the end, to 
add the non-valence contributions to the matrix elements of the 
electromagnetic current, all the prescriptions given the same results, 
colors solid line.

In the Fig.~\ref{qzeros},~(left),~the positions of the zeros for the 
charge electromagnetic form factor, 
is explored, and with the present model, the position of the 
zeros~, are linear with the rho meson bound squared mass; 
or with other words, the behavior of the zero positions of the 
charge electromagnetic form factors is given by~$q^2_{zero}\simeq 5 m^2_\rho$, 
its proportional the meson spin-1 bound state 
mass~(see the Fig.~\ref{qzeros}, the black solid line).

For the present work, the charge form factor zero appear around,~$Q^2=3.0~GeV^2$
for the experimental rho meson mass~$(770~MeV)$,~(see the 
also the Fig.~\ref{ffactorsg0g1}).
~Recently, the reference~\cite{Roberts2011},~with Schwinger-Dyson approach, 
found~$~5.0~GeV^2$. Late,~the based Schwinger-Dyson calculation,~\cite{Bhagwat2008}, 
have that zero about,~$3.8~GeV^2$. The "universal ratios"~\cite{Melo2016,Hiller92}, 
for the charge form factor,~with the experimental mass for rho 
meson,~$m_{\rho}=0.770~GeV$, give for the zero,~$q^2_{zero} \approx 3.6~GeV^2$, 
which is not far from the value predicted in this model.

\texttt{ }
\begin{figure*}[htb]  %%  FIG.4 
\includegraphics[scale=.27]{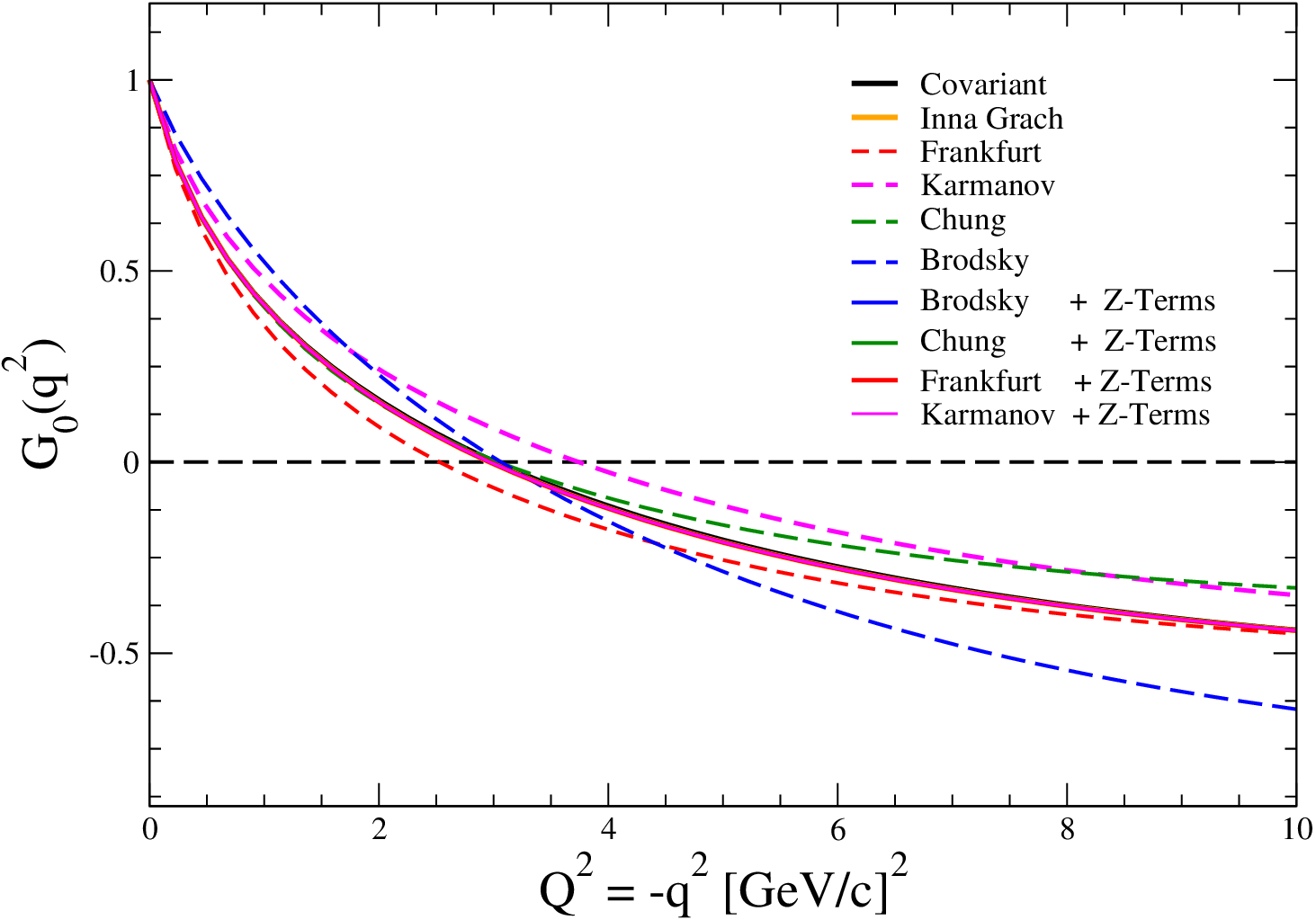}               
\includegraphics[scale=.27]{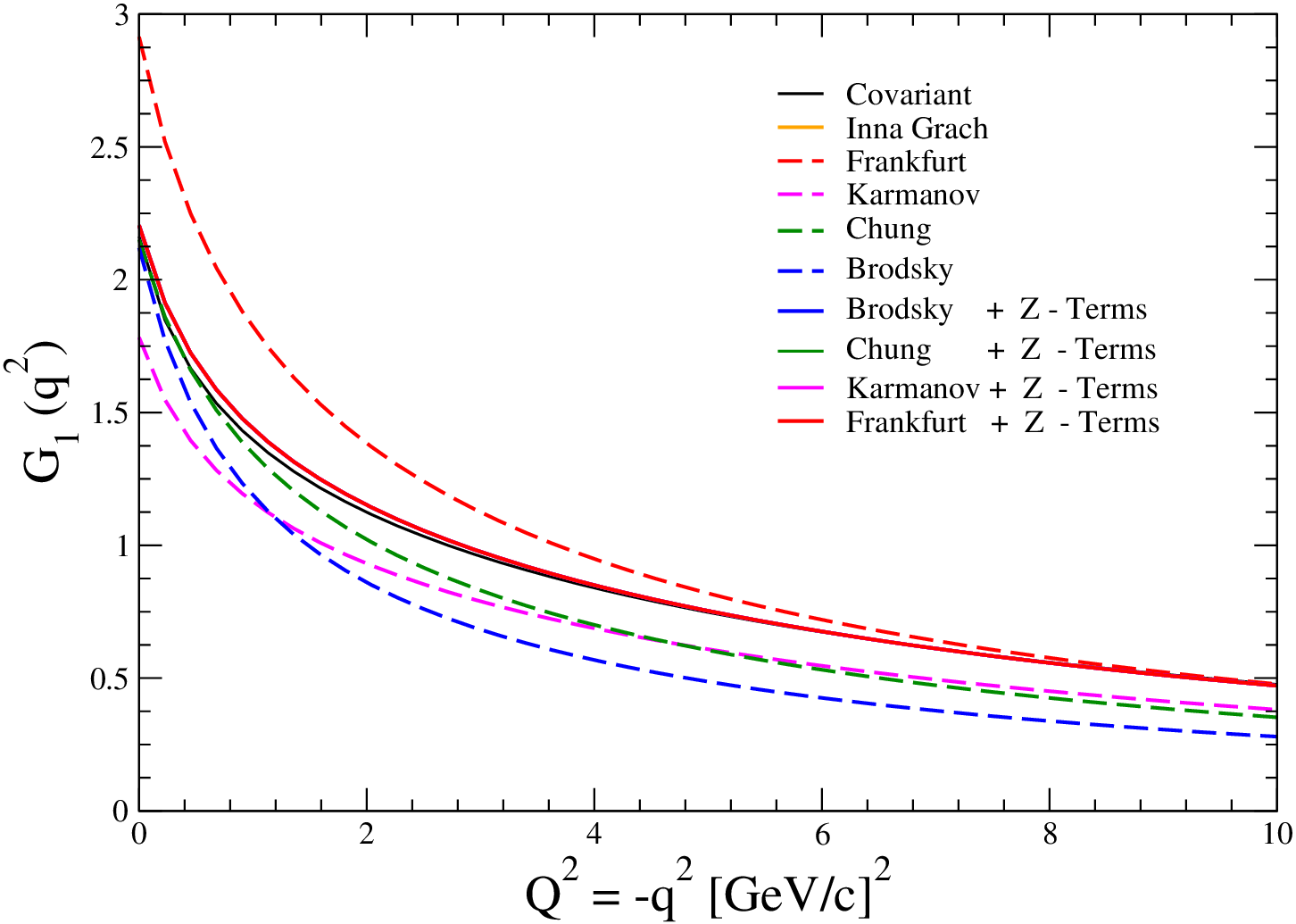}         
\caption{Charge form factor~(left) and magnetic form factor~(right) 
for the rho meson. The quark mass are $m_u=m_{\bar{d}}=0.430$~GeV, for the 
regulator mass~$m_R=3.0$~GeV, calculated 
with various prescriptions in the 
literature~\cite{Karmanov1996,Inna84,Chung88,Hiller92,Frankfurt93}.}
\label{ffactorsg0g1}
\end{figure*}

.
\vspace{0.3cm} 
\texttt{ }
\begin{figure*}[htbp]  %%  FIG.5
 \includegraphics[scale=.27]{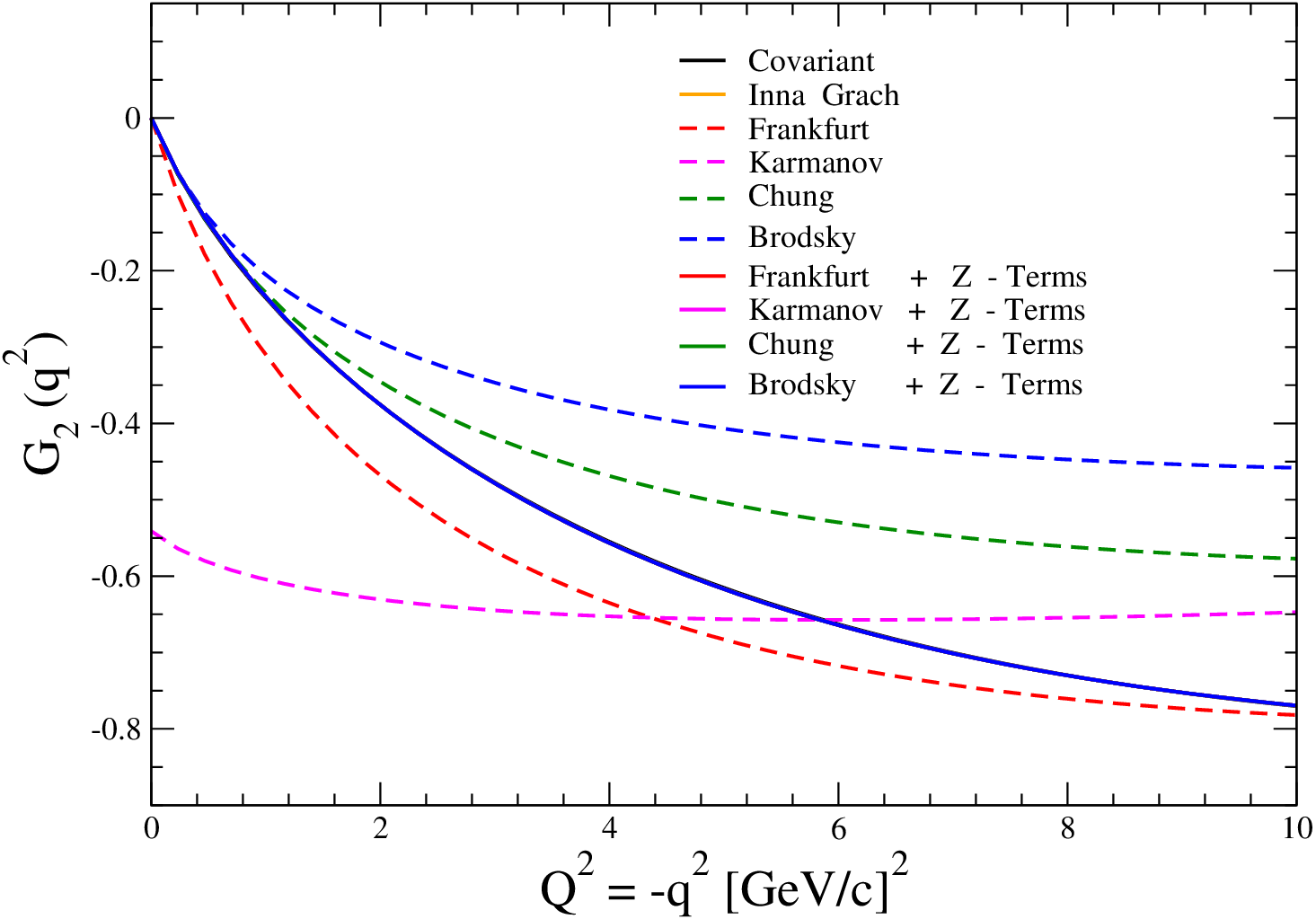}        
\caption{Quadrupole~$G_2(q^2)$ electromagnetic form factor, labels is the same from 
the Fig.\ref{ffactorsg0g1}
}
\label{ffactorg2}
\end{figure*}

\begin{figure*}[h!] %%tbp]  FIG.6
\begin{center}
\vspace{0.500020cm}
\includegraphics[scale=.28]{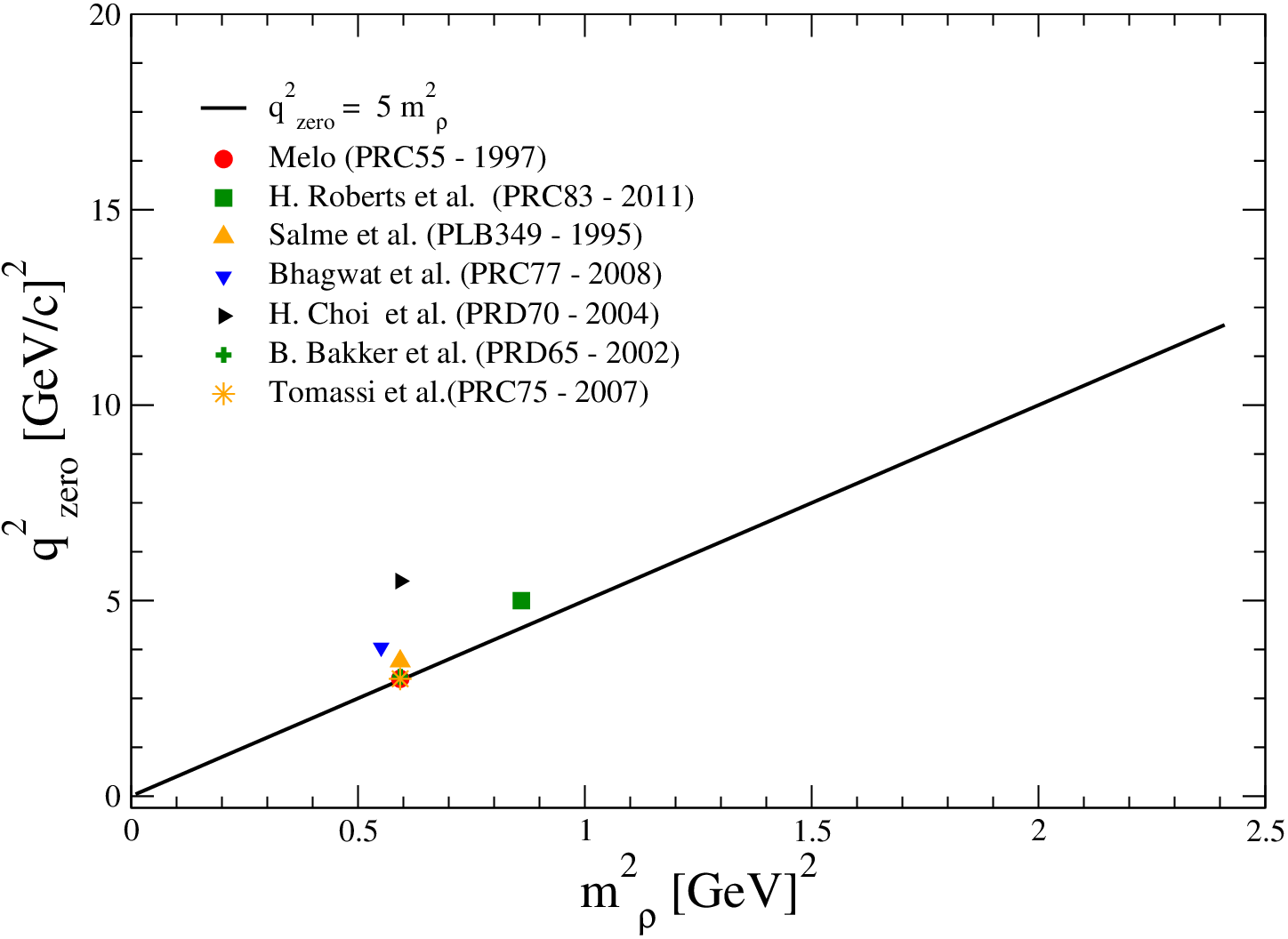}
\includegraphics[scale=0.28]{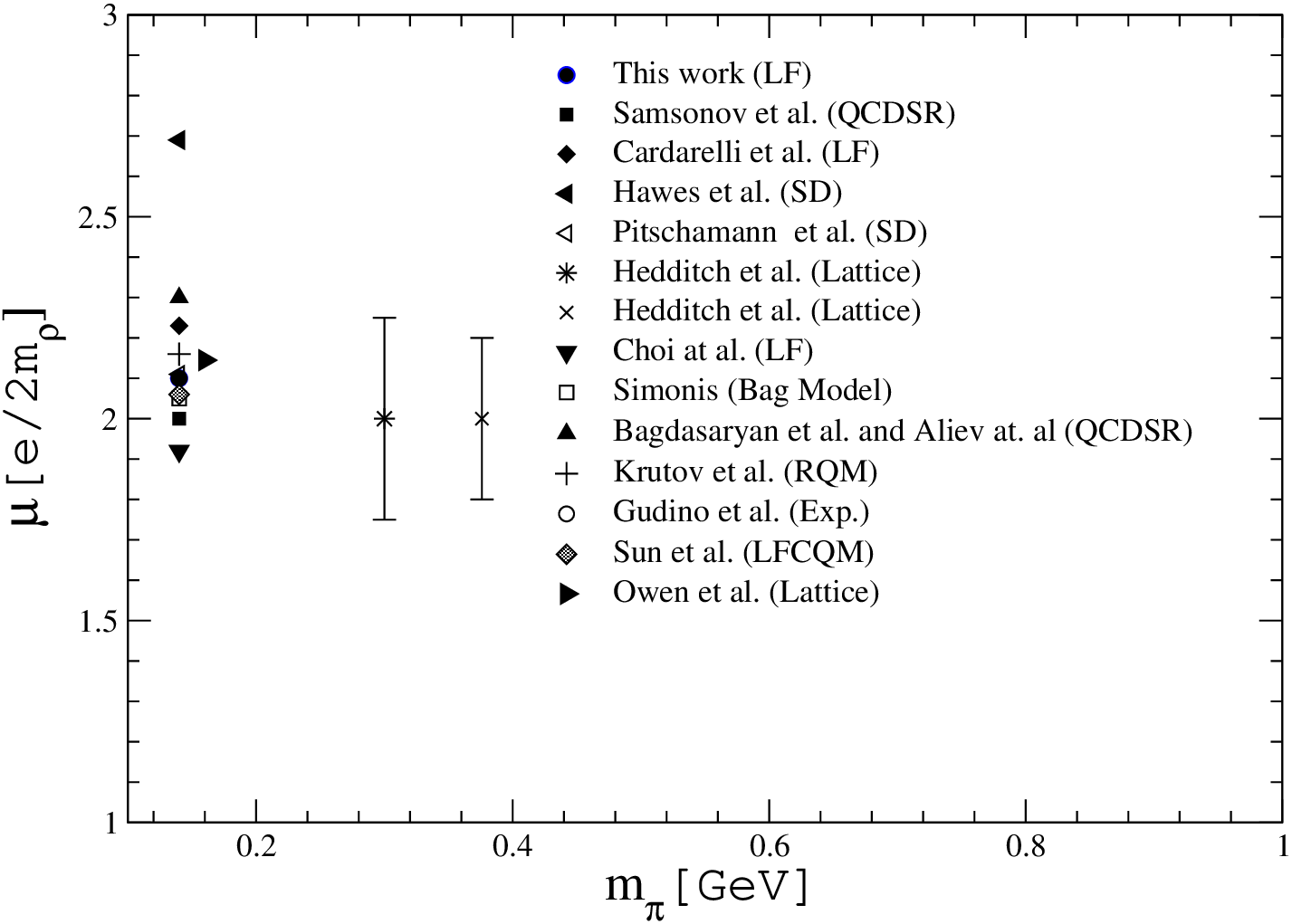}
\caption{(Left)~Charge electromagnetic form factor zero,~$G_0(q^2_{zero})=0$ in function of the 
rho meson mass, in the present model and compared with another's models~in the literature. 
(Right)~Magnetic moment for the rho meson compared with another 
models in the literature, 
~(QCDSR) QCD sum rules,~(LF)~Light-Front approach,
 (SD)~Schwinger-Dyson,~Lattice calculations,~Bag model, 
including with some experimental analysis. 
}
\label{qzeros}
\end{center}
\end{figure*}

The magnetic momentum calculated in the present work, is 
compared with others model from the literature in the 
Fig.~\ref{qzeros}~(right), in function of the pion mass, because in the same figure, 
the results from Lattice calculations are show~\cite{Hedditch2007}.

In the light-front approach, besides de valence 
components, we have non-valence contributions to the matrix elements 
of the electromagnetic current \cite{Pacheco97,Bakker2002,Pacheco99,Naus1999,Pacheco2002,Choi2014}. 
However, in the present work, independent of the prescription used
to extract the electromagnetic form factors and thus calculating some
observables, such as the decay constant, electromagnetic radius, magnetic and 
quadrupole momentum, we have obtained the same results;~for this, the relations between the 
matrix elements of the electromagnetic current at 
level of Dirac structure are fundamental,~Eq.~(\ref{relations2})~\cite{Pacheco2012}. 
With that relations, 
we arrival in Eqs.~(\ref{polizou}),~(\ref{hiller}) 
and,~(\ref{frederico}),~it is exactly the same equations utilized in the 
Inna prescription~\cite{Inna84}, in order to obtain the electromagnetic form factors for spin-1 particles, 
in case here, the rho meson. 
Also, with the Eq.(\ref{sum1}), the charge electromagnetic form factor, calculated in the 
instant form basis, not have any dependence with prescriptions, but, if calculated, with 
the light-front approach, the rotational symmetry is broken, and, 
after add the zero modes to the matrix elements of the electromagnetic current, the 
rotational symmetry is completely restored. The position of the zero for the 
charge electromagnetic form factor before the addition of the zero modes is around 
~$2.5~GeV^2$, after added zero modes, is around~$3.0~GeV^2$,~the same with the instant form basis 
calculations,~(see in the Fig.~\ref{ffactorsg0g1},~right panel).

Concluding, the present work, extends the previous works,
~\cite{Pacheco97,Pacheco2012},~for spin-1 particles with a light-front constituent quark model,
and show an unambiguous procedure to extract the electromagnetic form factor of the
plus component of the electromagnetic current,~free of the zero modes or pair terms 
contributions to the matrix elements of the electromagnetic current. 

A study of others components the electromagnetic current in order 
to extract electromagnetic form factors for particles of spin-1, 
and calculate the observables, it is in progress.

%%\vspace*{1ex}
{\it Acknowledgements.} 
This work was partially supported by the Funda\c c\~ao de Amparo \`a Pesquisa do Estado de
S\~ao Paulo,~FAPESP,~Brazil, Process 
\newline No.2015/16295-5,~and FAPESP Tem\'atico,Brazil,Process No.2017/05660-0. 
\newline Conselho Nacional 
de Desenvolvimento Cient\'ifico e Tecnol\'ogico~(CNPq), 
Brazil, grants Nos.~308025/2015-6 and 401322/2014-9. 
This work is part of the project INCT-FNA Proc. No. 464898/2014-5.


\begin{thebibliography}{99} 

\bibitem{Brodsky98}
S. J. Brodsky, H.-C. Pauli, and S. S. Pinsky, 
Phys. Rep. 301 (1998) 299.

\bibitem{Maris1999} P. Maris and P. C. Tandy, 
Phys. Rev. C 60 (1999) 055214.

\bibitem{Roberts1994} C. D. Roberts and A. G. Willians,  
Prog. Part. Nucl. Phys. 33 (1994) 477.

\bibitem{Roberts2000} C. D. Roberts and S. M. Schmidt, 
Prog. Part. Nucl. Phys. 45 (2000) S1.

\bibitem{Braguta2004} V. Braguta and A. I. Onishchenko, 
Phys. Lett.~B 591~(2004) 267.  %% -276

\bibitem{Braguta22004} V. V. Braguta and A. I. Onishchenko, 
Phys. Rev.~D 70~(2004) 033001.

\bibitem{Jaus2003} W.~Jaus,
%``Consistent treatment of spin 1 mesons in the light front quark model,''
  Phys.\ Rev.\ D 67~(2003)~094010.   %% doi:10.1103/PhysRevD.67.094010. 

\bibitem{Aliev2004} 
  T.~M.~Aliev and M.~Savci,
  %``Electromagnetic form factors of the rho meson in light cone QCD sum rules,''
  Phys.\ Rev.\ D 70~(2004)~094007. 
  
\bibitem{Savci2009} T. M. Aliev, A. \"{O}zpineci and M. Savci,  
Phys. Lett.~B 678~(2009) 470. 

\bibitem{Karmanov1996}V.~A.~Karmanov,
  %``On ambiguities of the spin-1 electromagnetic form-factors in light front dynamics,''
  Nucl.\ Phys.\ A~608~(1996) 316. %% doi:10.1016/0375-9474(96)00260-6

\bibitem{Braguta2008} V. Braguta, W. Lucha and D. Melikhov, 
Phys. Lett.~B 661~(2008) 354. 

\bibitem{Bakulev2009}
%% QCD sum rules with nonlocal condensates and the spacelike pion form factor
A.P. Bakulev, A.V. Pimikov, N.G. Stefanis, 
Phys. Rev.~D 79~(2009) 093010.

\bibitem{Teramond2008} S. J. Brodsky and G. F. de T\'eramond,
Phys. Rev. D 77~(2008) 056007.

\bibitem{Bijnens1998} J. Bijnens, P. Gosdzinky and P. Talavera,
Phys. Lett. ~B 429~(1998) 111.

\bibitem{Leitner2011} O. Leitner,  J.-F. Mathiot, N. A. Tsirova, 
Eur. Phys. J.~A 47~(2011) 17.

\bibitem{Karmanov2007} V. A. Karmanov, J.-F. Mathiot and V. A. Sminorv,
Phys. Rev. D 75 (2007) 045012.

\bibitem{Elmer2014}E. P. Biernat and  W. Schweiger, 
Phys. Rev.~C 89~(2014) 055205.

\bibitem{Perry90}
R. J. Perry, A. Harindranath, and K. G. Wilson, 
Phys. Rev. Lett.~65~(1990) 2959.  

\bibitem{Brodsky2004} S. J. Brodsky, J. R. Hiller, 
D. S. Hwang and V. A.Karmanov,  Phys. Rev. D 69~(2004) 076001.  

\bibitem{Pacheco97}J. P. B. C. de Melo and T. Frederico, 
Phys.~Rev.~C 55~(1997)~2043.

\bibitem{Choi2004} H.-M. Choi and C. R. Ji, 
Phys. Rev. D 70~(2004)  053015.

\bibitem{Bakker2002} B. L. G. Bakker and 
C. R. Ji, Phys. Rev. D 65~(2002) 116001.

\bibitem{Pacheco99}J. P. B. C. de Melo, H. W. L. Naus, 
and T. Frederico, Phys. Rev. C 59~(1999) 2278.
  
\bibitem{Naus1999}J. P. B. C. de Melo,  H. W. L. Naus, 
T. Frederico and P. U. Sauer, Nucl. Phys. A 660~(1999) 219.  
 
\bibitem{Li2018}
  Y.~Li, P.~Maris and J.~Vary,
  %``Frame dependence of form factors in light-front dynamics,''
  Phys.\ Rev.\ D 97 (2018) no.5,  054034. %% \ doi:10.1103/PhysRevD.97.054034. \ %%   [arXiv:1712.03467 [hep-ph]].
  
\bibitem{Pacheco2012}
J. P. B. C. de Melo and T. Frederico, Phys. Lett. B 708 (2012) 87.
 
\bibitem{Otoniel2012} 
  E.~O.~da Silva, J.~P.~B.~C.~de Melo, B.~El-Bennich and V.~S.~Filho,
  %``Pion and kaon elastic form factors in a refined light-front model,''
  Phys.\ Rev.\ C 86, 038202 (2012)~038202. %%  doi:10.1103/PhysRevC.86.038202
  
\bibitem{Maris97} P. Maris, C. D. Roberts, Phys. Rev. C 56 (1997) 3369.

\bibitem{Fabiano2007}F. P. Pereira, J. P. B. C. de Melo, T. Frederico, 
L. Tomio, Nucl. Phys. A 610~(2007) 610.
 
\bibitem{Pacheco2002} J.~P.~B.~C. de Melo, 
T. Frederico, E. Pace and G. Salm\`e, 
Nucl. Phys. A 707~(2002) 399; ibid. Braz. J. Phys. 33 (2003) 301.
 
\bibitem{Melikhov2002}   D.~Melikhov and S.~Simula,
  %``Electromagnetic form-factors in the light front formalism 
  %% and the Feynman triangle diagram: Spin 0 and spin 1 two fermion systems,''
 Phys.\ Rev.\ D 65~(2002)~094043. %%   doi:10.1103/PhysRevD.65.094043


\bibitem{Yabuzaki2015}G.~H.~Yabusaki, I.~Ahmed,~A.~Paracha,
~J.~P.~B.~C.~de~Melo and B.~Bennich, 
Phys.~Rev.~D 92,~(2015)~034017. 

\bibitem{Fanelli2016} 
  C.~Fanelli, E.~Pace, G.~Romanelli, G.~Salme and M.~Salmistraro,
  %``Pion Generalized Parton Distributions within a fully covariant constituent quark model,''
  Eur.\ Phys.\ J.\ C 76~no.~5,~(2016)~253. %%  doi:10.1140/epjc/s10052-016-4101-1
  
\bibitem{Pacheco2006} 
  J.~P.~B.~C.~de Melo, T.~Frederico, E.~Pace and G.~Salm\`e,
  Phys.~Lett.~B 581~(2004) 75; ibid.  Phys.\ Rev.\ D 73 (2006) 074013.
   
\bibitem{Adamuscin2007}
  C.~Adamuscin, G.~I.~Gakh and E.~Tomasi-Gustafsson,
  Phys.\ Rev.\  C 75 (2007) 065202;  A. Dbeyssi, E. Tomasi-Gustafsson, G.~I.~Gakh and C.~Adamuscin, 
  Phys.\ Rev.\  C 85 (2012) 048201.
 
\bibitem{Samsonov2003} A. Samsonov, JHEP 0312 (2003) 061.
  
\bibitem{Grigoryan2007}  H. R. Grigoryan,  Phys.  Lett.  B 662 (2008) 158.

\bibitem{Aliev2009}T. M. Aliev, A. Ozpineci and M. Savci, 
Phys. Lett.~B 678~ (2009) 470. 

\bibitem{Garcia2010}
  D. Garcia Gudino and G. Toledo Sanchez, 
Phys.  Rev. D 81~(2010) 073006.

\bibitem{Choi2011}
  H. M. Choi and C. R. Ji, Phys. Lett. B 696~(2011) 518.
    
 \bibitem{Pitschmann2013} M. Pitschmann, C.-Y. Seng, 
 M. J. Ramsey-Musolf, C. D. Roberts, S. M. Schmidt, D. J. Wilson, 
 Phys. Rev.~C 87~(2013) 015205.
 
\bibitem{Choi2014} 
H. M. Choi and C. R. Ji, Phys. Rev. D 89~(2014) 033011.
 
\bibitem{Melo2016} J.~P.~B.~C.~de Melo, C.~R.~Ji and T.~Frederico,
  %``The ρ -meson time-like form factors in sub-leading pQCD,''
  Phys.\ Lett.\ B 763,~(2016)~87. %%   doi:10.1016/j.physletb.2016.10.018
  
\bibitem{Krutov2016} 
  A.~F.~Krutov, R.~G.~Polezhaev and V.~E.~Troitsky,
  %``The radius of the rho meson determined from its decay constant,''
  Phys.\ Rev.\ D 93, (2016)~036007. 
  
\bibitem{Krutov2018} 
  A.~F.~Krutov, R.~G.~Polezhaev and V.~E.~Troitsky,
  %``Magnetic moment of the ρ meson in instant-form relativistic quantum mechanics,''
  Phys.\ Rev.\ D 97, (2018) 033007. %%  doi:10.1103/PhysRevD.97.033007

\bibitem{Sun2017}B.~D.~Sun and Y.~B.~Dong,
  %``$\rho$ meson unpolarized generalized parton distributions with a light-front constituent quark model,''
  Phys.\ Rev.\ D 96~(2017) 036019.   %%doi:10.1103/PhysRevD.96.036019 .
 
\bibitem{Sun2017v2} 
  B.~d.~Sun and Y.~b.~Dong,
  %``Deuteron electromagnetic form factors with the light-front approach,''
  Chin.\ Phys.\ C 41 (2017) 013102.   %%doi:10.1088/1674-1137/41/1/013102
  
 \bibitem{Pichowsky1999}
  F.~T.~Hawes and M.~A.~Pichowsky,
  %``Electromagnetic form-factors of light vector mesons,''
  Phys.\ Rev.\ C 59 (1999) 1743. %% \ doi:10.1103/PhysRevC.59.1743.
  
\bibitem{Hedditch2007}
  J.~N.~Hedditch, W.~Kamleh, B.~G.~Lasscock, D.~B.~Leinweber, A.~G.~Williams and J.~M.~Zanotti,
  %``Pseudoscalar and vector meson form-factors from lattice QCD,''
  Phys.\ Rev.\ D 75 (2007) 094504.    %%   doi:10.1103/PhysRevD.75.094504
  
\bibitem{Owen2015} 
  B.~Owen, W.~Kamleh, D.~Leinweber, B.~Menadue and S.~Mahbub,
  %``Light Meson Form Factors at near Physical Masses,''
  Phys.\ Rev.\ D 91 074503 (2015).   %% doi:10.1103/PhysRevD.91.074503
  
\bibitem{Shultz2015} 
  C.~J.~Shultz, J.~J.~Dudek and R.~G.~Edwards,
  %``Excited meson radiative transitions from lattice QCD using variationally optimized operators,''
  Phys.\ Rev.\ D 91 (2015) 114501.  %%  doi:10.1103/PhysRevD.91.114501
 
\bibitem{Blunden1996} 
  P.~G.~Blunden and G.~A.~Miller,
  %``Quark - meson coupling model for finite nuclei,''
  Phys.\ Rev.\ C 54~(1996) 359. \  doi:10.1103/PhysRevC.54.359.

\bibitem{Blunden1999} 
  P.~G.~Blunden, M.~Burkardt and G.~A.~Miller,
  %``Light front nuclear physics: Mean field theory for finite nuclei,''
  Phys.\ Rev.\ C 60~(1999) 055211. %%  \ doi:10.1103/PhysRevC.60.055211. 
  
\bibitem{Huber1998}
  G.~M.~Huber, G.~J.~Lolos and Z.~Papandreou,
  %``Probing the rho0 mass modification in the subthreshold region on He-3,''
  Phys.\ Rev.\ Lett.\ 80~(1998) 5285. %% \   doi:10.1103/PhysRevLett.80.5285
  
\bibitem{Huber2003} 
  G.~M.~Huber {\it et al.} [TAGX Collaboration],
  %``In-medium rho0 spectral function study via the H-2, He-3, C-12(gamma,pi+ pi-) reaction,''
  Phys.\ Rev.\ C 68 (2003) 065202. %%  \   doi:10.1103/PhysRevC.68.065202. 
  
\bibitem{Melo2014}
  J.~P.~B.~C.~de Melo, K.~Tsushima, B.~El-Bennich, E.~Rojas and T.~Frederico,
  %``Pion structure in the nuclear medium,''
  Phys.~Rev.~C 90~(2014)~035201.   %% \ doi:10.1103/PhysRevC.90.035201. 
 
 \bibitem{Melo2017}
  J.~P.~B.~C.~de Melo, K.~Tsushima and I.~Ahmed,
  %``In-Medium Pion Valence Distributions in a Light-Front Model,''
  Phys.\ Lett.\ B 766 (2017) 125.  %%   doi:10.1016/j.physletb.2017.01.004    [arXiv:1608.03858 [hep-ph]].
    
\bibitem{deAraujo2018} 
  W.~R.~B.~de Aráujo, J.~P.~B.~C.~de Melo and K.~Tsushima,
  %% ``Study of the in-medium nucleon electromagnetic form factors using a 
  %% light-front nucleon wave function combined with the quark-meson coupling model,''
  Nucl.\ Phys.\ A 970 (2018) 325.   %% doi:10.1016/j.nuclphysa.2017.12.005
  
\bibitem{deMelo2018b}
 J.~P.~B.~C.~de Melo and K.~Tsushima,
  %``$\rho$-meson properties in medium,''
  \newline (To appear Phys.Lett. B),~arXiv:1802.06096 [hep-ph].  
  
\bibitem{Cardarelli1995}
F. Cardarelli, I. L. Grach, I. M. Narodetsky, 
G. Salm\`e, S. Simula, Phys. Lett. B 349 (1995) 393.

\bibitem{Inna84} I. L. Grach and  L. A. Kondratyuk, Sov. J. Nucl. Phys.  38  (1984) 198; 
I. L. Grach, L. A. Kondratyuk, and M. Strikman, Phys.  Rev.  Lett.  62  (1989) 387.

\bibitem{Chung88} 
P. L. Chung, W. N. Polyzou, F. Coester and   B. D. Keister, Phys. Rev. C 37 (1988) 2000.

\bibitem{Hiller92} S. J. Brodsky and J. Hiller, Phys. Rev. D 46 (1992) 2141.

\bibitem{Frankfurt93} L. L. Frankfurt, T. Frederico, and M. Strikman, 
Phys. Rev. C 48
(1993) 2182.

\bibitem{Pacheco2004} J. P. B. C. de Melo and T. Frederico,  Braz.  J. Phys. 34 (2004) 881.

\bibitem{Pacheco98} J. P. B. C. de Melo, J. H. O. Sales, 
T. Frederico, and P. U. Sauer, Nucl. Phys. A 631~(1998) 574c.

\bibitem{Garcon2001} M.~Garcon and J.~W.~Van Orden,
  %``The Deuteron: Structure and form-factors,''
  Adv.\ Nucl.\ Phys.\  26~(2001) 293.  %%  doi:10.1007/0-306-47915-X_4.

\bibitem{Gilman2001}   R.~A.~Gilman and F.~Gross,
  %``Electromagnetic structure of the deuteron,''
  J.\ Phys.\ G 28 (2002) R37. %%   doi:10.1088/0954-3899/28/4/201

\bibitem{Bhagwat2008} M. S. Bhagwat and P. Maris, Phys. Rev. C 77 (2008) 025203.

\bibitem{Roberts2011} 
  H.~L.~L.~Roberts, A.~Bashir, L.~X.~Gutierrez-Guerrero, C.~D.~Roberts and D.~J.~Wilson,
  %``pi- and rho-mesons, and their diquark partners, from a contact interaction,''
  Phys.\ Rev.\ C 83~(2011)~065206.    %%  \ doi:10.1103/PhysRevC.83.065206
  
 \bibitem{PDG2018} The Review of Particle Physics (2018), 
 M. Tanabashi et al. (Particle Data Group), Phys. Rev. D 98 (2018) 030001.
%%  J. Beringer et al. (Particle Data Group), Phys. Rev. {\bf D 86} (2012)  010001.
 
\bibitem{Serrano2015} 
  M.~E.~Carrillo-Serrano, W.~Bentz, I.~C.~Cloët and A.~W.~Thomas,
  %``$\rho$ meson form factors in a confining Nambu–Jona-Lasinio model,''
  Phys.\ Rev.\ C 92~(2015)~015212. 

\bibitem{Gudino2015}   D.~Garc\'\i a  Gudiño and G.~Toledo S\'anchez,
  %``Determination of the magnetic dipole moment of the rho meson using four-pion electroproduction data,''
  Int.\ J.\ Mod.\ Phys.\ A 30~(2015)~1550114.
%%%``Determination of the magnetic dipole moment of the rho meson using 4 pion electroproduction data,''
ibid.,  Int.\ J.\ Mod.\ Phys.\ Conf.\ Ser.\  35~(2014) 1460463. 
  
\bibitem{Simonis2016} 
  V.~\v{S}imonis,   %``Magnetic properties of ground-state mesons,''
  Eur.\ Phys.\ J.\ A 52~(2016)~90. %%   doi:10.1140/epja/i2016-16090-5
  
\bibitem{Simonis2018}   V.~Simonis,
  %``Improved predictions for magnetic moments and M1 decay widths of heavy hadrons,''
  arXiv:1803.01809 [hep-ph].

 \end{thebibliography}
\end{document}